\documentclass[aps,twocolumn,floatfix,10pt]{revtex4-2}
\usepackage[utf8]{inputenc}
\usepackage[english]{babel}
\usepackage[T1]{fontenc}
\usepackage{amsmath,amssymb,amsfonts}
\usepackage{mathtools}

\usepackage{graphicx}
\usepackage[looser,scaled=1.05]{newtxtext} 
\usepackage[vvarbb,subscriptcorrection,scaled=1.05]{newtxmath} 

\usepackage{hyperref}
\hypersetup{
    colorlinks=true,
    linkcolor=blue,
    citecolor=blue,
    filecolor=blue,
    urlcolor=blue,
}

\usepackage[left=1.82cm,right=1.82cm,top=1.85cm,bottom=2.7cm]{geometry}
\addto\captionsenglish{}
\addto\captionsenglish{}

\begin{document}
\title{\vspace*{0.3cm}Recursive Magic State Distillation on the Surface Code}
\author{Jonathan E. Moussa}
\affiliation{Molecular Sciences Software Institute, Virginia Tech, Blacksburg, Virginia 24060, USA}
\email{godotalgorithm@gmail.com}

\begin{abstract}
\vspace*{0.6cm}
I reduce the cost to prepare magic states with\vspace*{1.4pt} lattice surgery operations on the surface code by using a
 recursive implementation of 15-to-1 magic state distillation.\vspace*{1.4pt} On a rotated surface code with distance $d$,
 $|T\rangle$ preparation requires a $d$-by-$3 d$ grid\vspace*{1.4pt} of data qubits for up to $15 d$ error correction cycles, and $|CCZ\rangle$
 preparation requires a $3 d$-by-$2 d$ grid for up to $10.5 d$ cycles.\vspace*{1.4pt} However, a significantly lower physical error
 threshold than that of the underlying surface code is \vspace*{1.4pt} required to match the error probability of the output
 magic state with the logical error rate \vspace*{1.4pt} of the output surface code at large code distances.
\vspace*{1.7cm}
\end{abstract}

\maketitle

\section{Introduction}
\vspace*{-0.25cm}

Universal quantum computation on physical qubits requires fault-tolerant quantum error correction to correct physical errors.
The effectiveness of this error correction is determined by the ratio of the physical error rate to the error threshold of an error correcting code.
The most effective code is often the surface code \cite{surface_code}, which has the highest known error threshold in addition to
 a convenient implementation on a square grid of physical qubits with physical gates between nearest neighbors.
Universal quantum computation on the surface code can be achieved by lattice surgery \cite{surgery1,surgery2,surgery3,surgery4} for logical Clifford gates
 and magic state distillation \cite{distillation1,distillation2} for logical non-Clifford gates.

Magic state distillation on a distance-$d$ surface code is known to have the same $O(d^3)$ spacetime cost as Clifford gates \cite{top_down1,top_down2}.
However, the ratio of cost prefactors between non-Clifford gates and Clifford gates was roughly one thousand in early implementations.
As a result, the design \cite{algorithm} and compilation \cite{compile} of quantum algorithms often uses the number of non-Clifford gates as a simple, representative cost metric.
Improved implementations of magic state distillation on the surface code \cite{distillation3,distillation4,distillation5,distillation6,distillation7} have reduced the cost prefactor for a limited range of distances,
 but they have not yet established a clear implementation or cost prefactor at large code distances.
Alternative designs for magic state distillation have reduced the ratio of input magic states to output magic states \cite{block1,block2,block3},
 but they have not yet provided a clear implementation or cost prefactor on the surface code.
Thus, the number of ideas for improving implementations of magic state distillation on the surface code has been growing,
 but the understanding of their impact on asymptotic cost has been stagnant.

In this paper, I reduce the prefactor in the $O(d^3)$ spacetime cost of magic state distillation on a distance-$d$ surface code
 with a recursive implementation of 15-to-1 distillation of $|T\rangle$ in a spatial footprint of three surface code tiles.
I first prepare nine distance-$\lfloor d/3 \rfloor$ magic states simultaneously and pack them before preparing six more magic states simultaneously in the remaining space.
I then encode these magic states in the quantum Reed-Muller code using lattice surgery stabilizer measurements to detect errors in both magic states and lattice surgery operations.
I encode the stabilizer measurements in a classical error detecting code \cite{temporal_encoding} to satisfy a set of fault tolerance requirements.
To reduce the time cost, I carefully choose stabilizer generators and their measurement schedule to minimize movement and maximize concurrency.
I use the same ideas to implement a terminal $8 |T\rangle$-to-$|CCZ\rangle$ distillation step \cite{ccz} at distance $d$.
In this implementation, it is more efficient to teleport a $CCZ$ gate from one $|CCZ\rangle$ rather than four $|T\rangle$.
Finally, I perform a preliminary error analysis of this distillation process, which identifies a physical error threshold that is much smaller than the threshold of the underlying surface code.
This distillation process is still functional above its threshold, but it requires an operational surface code distance that is larger than the output surface code distance to balance error probabilities.

\section{Framework}
\vspace*{-0.25cm}

I mainly consider surface codes from the macroscopic perspective of topological objects rather than the microscopic perspective of physical qubits.
I only use the microscopic perspective for modeling costs and errors.
I assume that a square grid of distance-$d$ rotated surface codes is implemented on contiguous $d$-by-$d$ tiles of a square grid of physical qubits that persistently store data.
In typical implementations, there is also an interleaved square grid of physical qubits that are used as temporary ancilla to measure the stabilizer generators of the surface codes.
Together, the data and ancilla qubits form a square grid of physical qubits rotated by $45^{\circ}$,
 and stabilizer measurements are implemented using physical gates between nearest neighbors on this grid.
In typical cost models, space is measured in the number of data qubits, time is measured in the number of error correction cycles, and spacetime is measured as their product, qubitcycles.

\begin{figure}
\includegraphics{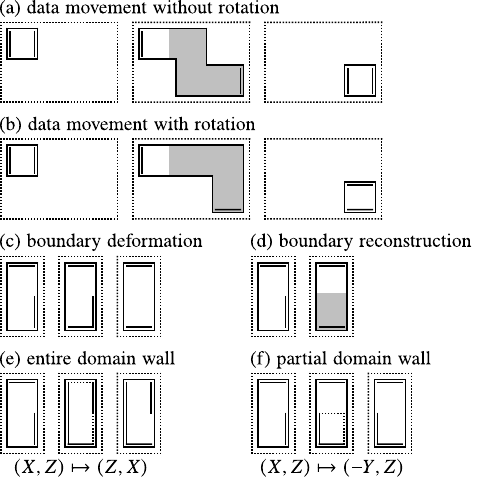}
\caption{Diagrammatic examples of deformation operations on surface code patches.
(a) A surface code tile can be moved by an intermediate deformation into a patch connecting its initial and final location, and (b) it can also be rotated if the patch contains an odd number of turns.
Boundaries of a surface code patch can be deformed if they are connected by (c) timelike or (d) spacelike boundaries that preserves the code distance.
A domain wall can cover (e) an entire surface code patch and switch all boundary types or
 (f) part of a surface code patch if twist defects in the interior of the patch partition it into regions of preserved and switched boundary types.
Each type of domain wall applies a different map to the logical Pauli operators of the patch.
}\label{fig_notation}
\end{figure}

Adapting previous work \cite{surgery3,distillation4}, I visualize lattice surgery operations on surface code patches as sequences of two-dimensional images in natural reading order,
 each depicting a spatial layout for a time interval spanning some number of error correction cycles.
I provide basic examples of this diagrammatic notation in Fig.\@ \ref{fig_notation}.
Alternatively, such sequences can be assembled into three-dimensional spacetime structures and visualized as graphical projections \cite{surgery2,surgery4,distillation3}.
Each computational step in the sequence is bordered by a dotted line depicting the static boundary of its computational region.
The rectilinear shapes bordered by solid lines depict the surface code patches in each time interval, and patches are decorated to define their structure and connectivity.
Also, some surface code patches may be labeled by a logical entity such as a state, operation, or measurement outcome.

Single lines denote timelike rough boundaries of surface code patches, and double lines denote timelike smooth boundaries.
By default, surface code patches are topologically connected to the patches at the same location in the previous time interval.
Otherwise, shaded regions of a surface code patch denote spacelike boundaries at the beginning of the time interval that separate timelike boundaries of the opposite type (rough or smooth).
Physical qubits are initialized to $|0\rangle$ to form an initial rough boundary and to $|+\rangle$ to form an initial smooth boundary.
In regions with no connection to a patch in the next time interval, patches have a spacelike boundary at the end of the time interval that also separates timelike boundaries of the opposite type.
Physical qubits are measured in the Pauli $Z$ basis to form a terminal rough boundary and in the Pauli $X$ basis to form a terminal smooth boundary.
Except for physical injection of $|T\rangle$, I initialize new surface code patches with spacelike smooth boundaries, which corresponds to logical preparation of $|0\rangle$.
The spacelike boundary that terminates a surface code patch is irrelevant unless the corresponding logical measurement outcome is needed as output or to condition a subsequent operation.
The logical measurement basis is specified in context when relevant, with a rough boundary for $X$ and a smooth boundary for $Z$.

Regions of a surface code patch surrounded by solid and dotted lines denote spacelike domain walls at the beginning of the time interval.
The simplest case is a domain wall that covers an entire surface code patch as in Fig.\@ \ref{fig_notation}e.
Solid lines denote timelike boundaries changing from rough to smooth at the domain wall, and dotted lines denote timelike boundaries changing from smooth to rough.
An entire domain wall is implemented by applying Hadamard $H$ gates to all physical qubits in the patch.
This operation acts like a logical $H$ gate and switches conjugate pairs of logical Pauli operators between $X$ and $Z$.
It also switches the Pauli type of stabilizer generators between $X$ and $Z$, which requires a surface code deformation to realign the stabilizers with the patch before the domain wall.

The topological structure of a surface code operation is delineated by one-dimensional twist defects \cite{surface_twist,diagonal_s_gate}.
In general, the two-dimensional spacetime boundary of a surface code operation is covered by topologically distinct regions that are separated by twist defects.
For an idle surface code tile, each side is a topologically distinct timelike rough or smooth boundary separated by timelike twist defects in each corner.
If twist defects remain on the spacetime boundary, then topologically distinct regions have a single boundary type.
However, twist defects can also pass through the interior of the spacetime region as edges of a partial domain wall.
Partial domain walls are surfaces within the spacetime region that either terminate at a twist defect in the interior or at a boundary between rough and smooth regions that are not topologically distinct.
Thus, topologically distinct surface region can include both rough and smooth boundary types because of intersections with domain walls.
Globally, twist defects form closed loops in spacetime, although their closure and connectivity may not be clear from a single surface code operation within a larger quantum computation.

Here, I only consider spacelike partial domain walls bordered by a spacelike twist defect, as shown in Fig.\@ \ref{fig_notation}f.
The spacelike twist defect is denoted by a dotted line through the patch at the internal boundary of the partial domain wall.
The example in Fig.\@ \ref{fig_notation}f depicts an initial timelike defect on the right side that is connected to a final timelike defect on the left side through the spacelike defect.
With a smooth boundary adjacent to both the initial timelike twist defect and the partial domain wall, the logical effect of this operation is either an $S$ or $S^\dag$ gate.
This is an arbitrary choice of error correction convention, and I choose the logical map of $S$, $(X,Z) \mapsto (-Y,Z)$ for $Y = i XZ$.

A useful property of topological error correcting codes such as the surface code is that their code distance is related to a geometric distance in space.
Similarly, the fault tolerance distance of a surface code operation is related to a geometric distance in spacetime.
On the rotated surface code, there are two relevant geometric distances, one for logical $X$ and $Z$ error strings and another for logical $Y$ error strings.
These geometric distances are determined by patterns of data and measurement errors that are not locally detectable by the stabilizer measurements of the code.
For $X$ and $Z$ types, the distance between two points is the maximum of their horizontal and vertical lattice separation on the underlying grid of data qubits plus the number of error correction cycles separating them.
For $Y$ type, the distance between two points is the sum of their horizontal and vertical lattice separation plus twice the number of error correction cycles separating them.
In both cases, surfaces of equal distance from a point are squares in space that extend to octahedra in spacetime.
However, the size and alignment of these surfaces are different, and $Y$ surfaces are contained inside $X$ and $Z$ surfaces for the same origin and distance.
The reason for this behavior is that $Y$ error strings can only be extended in most directions by pairs of distinct $X$ and $Z$ errors in data or measurement.
Specifically, $Y$ error strings can only be extended by physical $Y$ errors horizontally and vertically in space.

The code distance of a surface code patch is the minimum geometric distance of error strings through the patch that connect topologically distinct pairs of boundaries or twist defects.
Similarly, the fault tolerance distance of a surface code operation is the minimum distance of strings through its active spacetime region that connect distinct pairs of boundaries or twist defects.
$Z$ error strings connect to rough boundaries, $X$ error strings connect to smooth boundaries, and $Y$ error strings connect to twist defects.
An error string crossing a domain wall switches its type between $X$ and $Z$, but $Y$ error strings are invariant to domain wall crossings.
Error strings thus connect same-type boundaries when crossing an even number of domain walls and opposite-type boundaries when crossing an odd number of domain walls.
Each distinct pair of boundaries or twist defects connected by an error string corresponds to a topologically distinct logical error.

The surface code operations discussed in this paper are restricted to rectilinear spacetime twist defects.
The smallest surface code patch with a code distance of $d$ is then a $d$-by-$d$ tile with a timelike twist defect in each corner.
Also, each time interval requires $d$ error correction cycles to achieve a fault tolerance distance of $d$ on a surface code patch of distance $d$.
Unrestricted surface code designs can reduce the space of codes \cite{twisted_surface} and the time of operations \cite{diagonal_s_gate} but increase the complexity of implementation.

\subsection{Lattice surgery}
\vspace*{-0.25cm}

\begin{figure}
\includegraphics{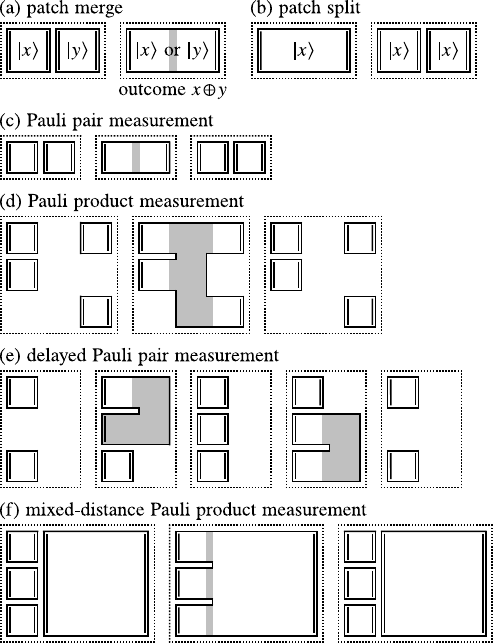}
\caption{Diagrammatic examples of lattice surgery operations on surface code patches.
Patches can be (a) merged along a smooth boundary to measure the product of their logical Pauli $Z$ operators or (b) split to copy their state in the eigenbasis of $Z$, $|x\rangle$.
The output state of a merge depends on a choice of error correction conventions.
(c) A smooth merge and split can be combined into a projective $Z \otimes Z$ measurement if a Pauli $X$ correction is applied to the patch that is altered by the merge after outcome 1.
(d) A smooth merge and split with an intermediate connection between multiple patches projectively measures the product of their $Z$ operators.
(e) Intermediate ancilla patches can also be used to delay part of a Pauli product measurement.
(f) These Pauli measurement operations generalize to patches of varying code distance.
Similarly, merges and splits along a rough boundary measure products of $X$ and act on its eigenbasis, $H|x\rangle$ (i.e.\@ $|+\rangle = H|0\rangle$ and $|-\rangle = H|1\rangle$).
}\label{fig_surgery}
\end{figure}

Lattice surgery operations change the topology of surface code patches to perform logical Pauli measurements.
The simplest such operation is the merging and splitting of two patches \cite{surgery1}, which can be combined with ancilla patches to measure arbitrary products of Pauli operators \cite{surgery3}.
In addition to these established operations, I also consider splitting and merging between surface code patches that have different code distances.
I summarize these elementary lattice surgery operations in Fig.\@ \ref{fig_surgery}.
All of these operations can be understood through changes to logical Pauli operators and surface code stabilizers.

Throughout the paper, I label all measurement outcomes as either $0$ or $1$.
This is the standard convention for measurements in the computational basis, while Pauli operator measurements are usually labeled by their eigenvalues of $-1$ and $1$.
Instead, I label the measurement outcomes of any Pauli operator $P$ as 0 and 1, corresponding to the eigenvalues of the projection operator $(I-P)/2$.
This notation aligns the outcome labels of Pauli $Z$ measurements and computational basis measurements.

To describe the smooth merging and splitting of two surface code patches in Fig.\@ \ref{fig_surgery}c, I define two alternative conjugate pairs of logical Pauli operators,
\begin{align} \label{pair_logical}
 X'_1 = X_1 X_2, \ \ \ & Z'_1 = Z_1, \notag \\
 \mathrm{and} \ \ \ X'_2 = X_2, \ \ \ & Z'_2 = Z_1 Z_2,
\end{align}
 where $X_1$ and $Z_1$ are logical Pauli operators of the left patch and $X_2$ and $Z_2$ are logical Pauli operators of the right patch.
Specifically, I choose $X'_1$ to be a physical $X$ string connecting smooth boundaries of the merged patches and $Z'_2$ to be a physical $Z$ string that forms a closed loop.
$X'_1$ and $Z'_1$ are then logical operators of both the split and merged patches, while $X'_2$ and $Z'_2$ are logical operators of only the split patches.
Merging moves the $Z'_2$ operator into the stabilizer group, which causes it to be measured.
If the $Z'_2$ stabilizer is measured as 1, $X'_2$ is applied to stabilize the merged patches, and it is applied again to the final split patch to implement the projective measurement of $Z'_2$.
Eq.\@ (\ref{pair_logical}) corresponds to the choice of assigning the state of the left patch to the merged patches, while $X'_2 = X_1$ and $Z'_1 = Z_2$ correspond to the alternative choice.
This choice is irrelevant to merges and splits that are combined into projective measurements.
Rough merging and splitting has an equivalent structure with $X$'s and $Z$'s switched.

Pauli product measurements based on smooth merging and splitting of multiple surface code patches such as in Fig.\@ \ref{fig_surgery}d have a similar description.
I define $n$ alternative conjugate pairs of logical Pauli operators for $n$ patches,
\begin{align}
 X'_1 = X_1 X_n, \ \ \ & Z'_1 = Z_1, \notag \\
 & \vdots \notag \\
 X'_{n-1} = X_{n-1} X_n, \ \ \ & Z'_{n-1} = Z_{n-1}, \notag \\ 
 \mathrm{and} \ \ \ X'_n = X_n, \ \ \ & Z'_n = Z_1 \cdots Z_n,
\end{align}
 where $X_m$ and $Z_m$ are logical Pauli operators of the $m$th patch.
Some operators are transformed by the preparation of the connecting ancilla patch during the merge.
$X'_m$ are extended by the initial spacelike smooth boundary of the ancilla patch into physical $X$ strings connecting smooth boundaries of patch $m$ and $n$.
$Z'_n$ becomes a stabilizer of the merged patch because the rough boundaries of the ancilla patch connect all $Z_m$ into a physical $Z$ string that forms a closed loop.
Otherwise, the descriptions of the merge, split, and projective $Z'_n$ measurement are the same as in the two-patch case.
The $X'_m$ and $Z'_m$ operators for $m < n$ remain logical operators while $Z'_n$ is measured during its temporary inclusion into the stabilizer group.

A Pauli product measurement can also be delayed as shown in Fig.\@ \ref{fig_surgery}e.
The delay can be described as two Pauli measurements that include an ancilla patch prepared in a complementary state.
The complementary state is $|+\rangle$ for $Z$ measurements and $|0\rangle$ for $X$ measurements.
Each intermediate measurement has a uniformly random outcome, and the product of their outcomes is the outcome of the delayed measurement.

A merge and split between surface code patches with different code distances such as in Fig.\@ \ref{fig_surgery}f has the same logical structure as in the previous cases.
The overall fault tolerance distance of this operation cannot exceed the lowest code distance $d$, but errors can be partitioned into topologically distinct strings of different minimum length.
Additional layers of error correction or error detection are required for shorter strings to preserve the higher overall fault tolerance distance in mixed-distance surface code operations.
Also, the size and shape of ancilla patches must be chosen carefully to preserve the minimum length of error strings from the higher-distance codes before and after a lattice surgery operation.

The number of error correction cycles during the measurement operations in Fig.\@ \ref{fig_surgery} determines the length of timelike error strings that cause measurement errors.
This number is usually equal to the lowest code distance to balance the length of the timelike and spacelike error strings that limit the fault tolerance distance.
More error correction cycles can be used in special cases when measurement errors must be suppressed more than data errors.
Fewer error correction cycles can be used if the measurement of multiple, commuting Pauli products is encoded in a classical error correcting code \cite{temporal_encoding}.
Such encodings increase the total number of measurements while reducing the time per measurement, which can reduce the overall time cost in some cases.
In Sec.\@ \ref{design_section}, I use encoded Pauli product measurements to satisfy fault tolerance requirements for $|T\rangle$ distillation circuits.
However, this encoding is used to detect measurement errors without further increasing the speed of measurements.
It could be beneficial to increase both measurement speed and code distance in these distillation circuits, but this expanded design space will need a more general set of fault tolerance requirements.

\subsection{Single-qubit Clifford gates}
\vspace*{-0.25cm}

\begin{figure}
\includegraphics{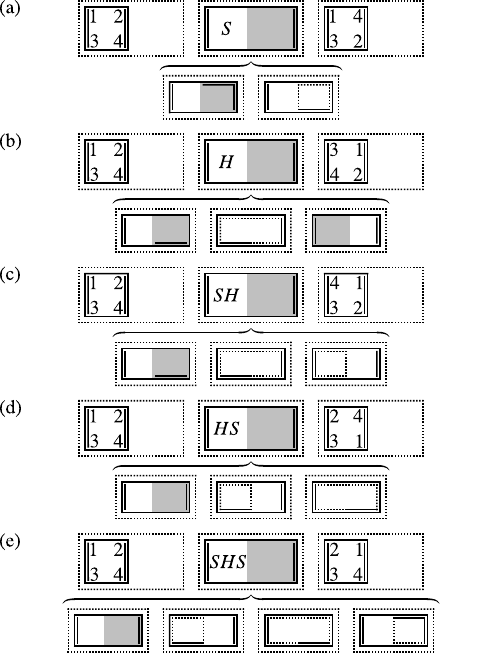}
\caption{Single-qubit Clifford gates for all nontrivial elements of the Clifford group up to global phases and Pauli corrections: (a) $S$, (b) $H$, (c) $SH$, (d) $HS$, and (e) $SHS$.
These implementations have a common spatial footprint and preserve the initial position and orientation of the surface code patch.
The timelike twist defects at the corners of the initial and final patches are labeled to visualize their distinct permutation by each Clifford gate.
}\label{fig_clifford}
\end{figure}

I consider the cost of logical Clifford gates as a reference for the cost of logical non-Clifford gates, starting with single-qubit gates.
These costs vary depending on design constraints and scheduling requirements.
I first consider the most expensive case, when Clifford gates must preserve the location and orientation of the surface code patch and have a common location for their ancilla patch.
From this reference point, a relaxation of scheduling requirements can reduce gate costs, but this reduction may be offset by increased data movement or less efficient packing of data.
I also consider the relaxation of design constraints.

To save some spacetime costs, I implement gates for every non-trivial element of the single-qubit Clifford group rather than generating elements from $H$ and $S$ gates.
Elements of the Pauli subgroup generated by $Z = SS$ and $X = HSSH$ are transversal logical gates that reduce to logical strings by the stabilizer group of the surface code.
The global phase subgroup generated by $e^{i\pi/4} I = HSHSHS$ is irrelevant to quantum computation.
Non-trivial representative elements of the remaining quotient group relative to these subgroups are $S$, $H$, $SH$, $HS$, and $SHS$.

I implement the single-qubit Clifford gates in Fig.\@ \ref{fig_clifford} using the surface code deformation operations from Fig.\@ \ref{fig_notation}.
There are two ways to understand the design of these gates.
Entire domain walls act as $H$ gates, partial domain walls act as $S$ gates, and the remaining deformations return the patch to its initial configuration.
Alternatively, Clifford gates are permutations of the timelike twist defects at the corners of a surface code patch \cite{surface_twist,diagonal_s_gate}.
Each step moves the twist defects through a surface code patch as part of a permutation while preserving the code distance by preserving the geometric distance between twist defects.
There are multiple permutations that correspond to each Clifford gate and multiple ways to implement each permutation, thus these implementations are not unique.
The $S$ gate has the fastest implementation here because it can be expressed as a permutation of the two twist defects that are adjacent to the ancilla patch.
The $SHS$ gate is also a permutation of two twist defects, but it has the slowest implementation because of poor alignment with the ancilla patch.

Up to a Pauli correction, the permutation of twist defects can be used as a heuristic for the transformation of logical Pauli operators.
The connectivity of $Y$ strings between twist defects is preserved by Clifford gates, and a permutation of twist defects corresponds to a spatial transformation of an attached string operator.
These $Y$ strings can be considered as a product of $X$ and $Z$ strings with the same connectivity, but each $X$ and $Z$ string can be in either the stabilizer group or the logical group.
If two twist defects are on both sides of a common rough boundary, then the $Z$ string is absorbed into the stabilizer group leaving a logical $X$ string.
Similarly, two twist defects adjacent to a common smooth boundary absorb the $X$ string into the stabilizer group and leave behind a logical $Z$ string.

Clifford gates on surface code patches can be faster if their scheduling requirements are relaxed.
It is well known that an entire domain wall as in Fig.\@ \ref{fig_notation}e is a transversal $H$ gate on a surface code tile that rotates it by $90^\circ$.
Also, the $SHS$ gate in Fig.\@ \ref{fig_clifford} can be as fast as the $S$ gate if its ancilla patch is adjacent to a rough boundary rather than a smooth boundary.
These two changes reduce the average time cost of implementing a non-trivial single-qubit Clifford group element from $3d$ to $1.6d$ error correction cycles.
However, the variable orientation of a patch or gate may increase costs in some cases by requiring more free tiles or complicating the schedule of logical operations.
These are tradeoffs between low-level considerations of gate design and high-level considerations of architecture, compilation, and scheduling that are beyond the scope of this paper.

The $H$ gate in Fig.\@ \ref{fig_clifford}b is an adaptation of its original lattice surgery implementation \cite{surgery1} with an additional movement step to return the patch to its initial position.
The $S$ gate in Fig.\@ \ref{fig_clifford}a is the implementation of Bomb\'{i}n \textit{et al.} \cite{surgery4} rather than the faster $S$ gate proposed by Gidney \cite{diagonal_s_gate}.
Gidney's $S$ gate only needs $\approx 1.5d$ error correction cycles instead of $2d$, but it requires the implementation of diagonal spacelike twist defects.
Twist defects also can be moved through the interior of a surface code patch without a spacelike alignment \cite{surface_twist}, which can further increase the speed of an $S$ gate to $d$ error correction cycles.

While unconstrained twist defects enable faster Clifford gates, they are more complicated to implement than rectilinear twist defects.
As discussed in previous work \cite{surgery4}, a vertical or horizontal spacelike twist defect is implemented by measuring physical qubits along the defect in the $Y$ basis.
This operation fuses pairs of weight-4 stabilizer generators that overlap with the defect into new weight-4 stabilizers of mixed Pauli type ($X$ and $Z$).
The partial domain wall is implemented by applying Hadamard $H$ gates to all physical qubits in the wall.
This transforms the fused stabilizer generators into a pure Pauli type, and a surface code deformation is also required to realign the stabilizer generators with the patch before the domain wall.
A diagonal spacelike twist defect complicates the implementation near the defect \cite{diagonal_s_gate}, and general twist defects \cite{surface_twist} require weight-5 stabilizer measurements.

\subsection{Multi-qubit Clifford gates}
\vspace*{-0.25cm}

\begin{figure}
\includegraphics{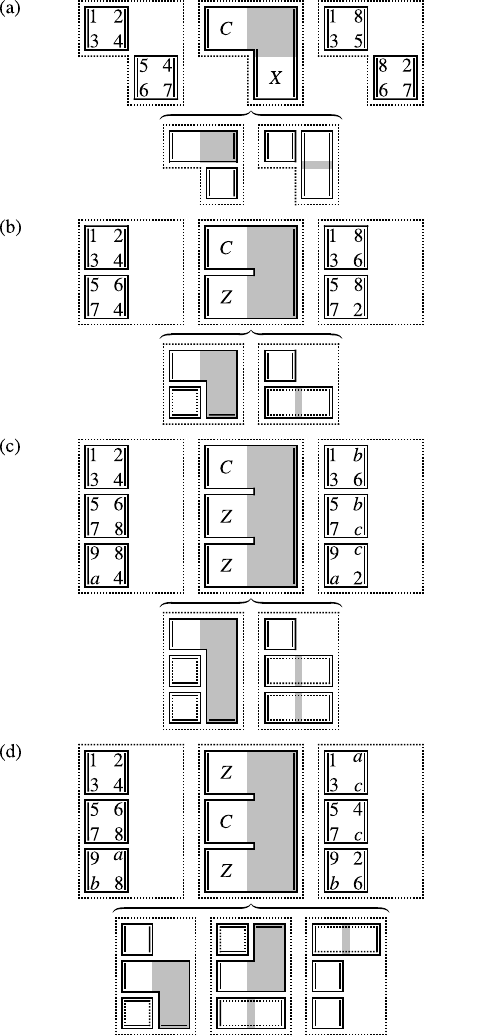}
\caption{Multi-qubit Clifford gates (a) $CX$, (b) $CZ$, (c) $CZ^2$ with targets on one side, and (d) $CZ^2$ with targets on two sides.
Pairs of connected twist defects are labeled before and after each gate.
}\label{fig_multi}
\end{figure}

Operations from the multi-qubit Clifford group are usually generated from a limited set of few-qubit Clifford gates in addition to the single-qubit Clifford gates.
The $CX$ gate in Fig.\@ \ref{fig_multi}a is the few-qubit Clifford gate from the original lattice surgery implementation \cite{surgery1}.
Later revisions of lattice surgery introduced multi-target $CX$ gates \cite{twist_surgery}, $CZ$ gates \cite{surgery2}, and a variety of $CX$ gate implementations \cite{zx_calculus}.
For consistency with the single-qubit Clifford gates in Fig.\@ \ref{fig_clifford}, I construct single-target and multi-target $CZ$ gates in Fig.\@ \ref{fig_multi} with ancilla patches adjacent to smooth boundaries.
This allows for a tight packing of surface code tiles with adjacent rough boundaries while computing.
A two-sided multi-target $CZ$ gate can be as fast as the one-sided version if it is implemented as two simultaneous one-sided gates on both sides of the control patch, which uses more space.
All of these $CZ$ gates are effectively $CX$ gates in between two transversal $H$ gates on the target patch.

The gates in Fig.\@ \ref{fig_multi} act between nearby surface code patches, but they are straightforward to adapt to control and target patches that are more separated and to more targets.
The speed of the gates is preserved, but the size of the computational region will increase when extending the ancilla patch between separated control and target patches.
Alternatively, controls and targets could be moved closer together prior to a multi-qubit gate.
One of the challenges of compilation and scheduling is to minimize the overall spacetime cost by determining when additional data movement is more efficient than multi-qubit gates between distant qubits.

Unlike Fig.\@ \ref{fig_clifford}, the gates in Fig.\@ \ref{fig_multi} do not utilize the entire computational region in every step.
It is simple and convenient to quantify costs and schedule operations as specific computational regions occupied for a specific interval of time, but it is not optimal.
If an operation is described more precisely by only its active spacetime region, then its effective spacetime cost may be reduced and a schedule of operations may be packed more tightly.
However, this extra precision may also increase the difficulty of computation and scheduling.
In this paper, I account for the cost of an operation simply by the size of its computational region and the duration of its time interval.

\begin{figure*}
\includegraphics{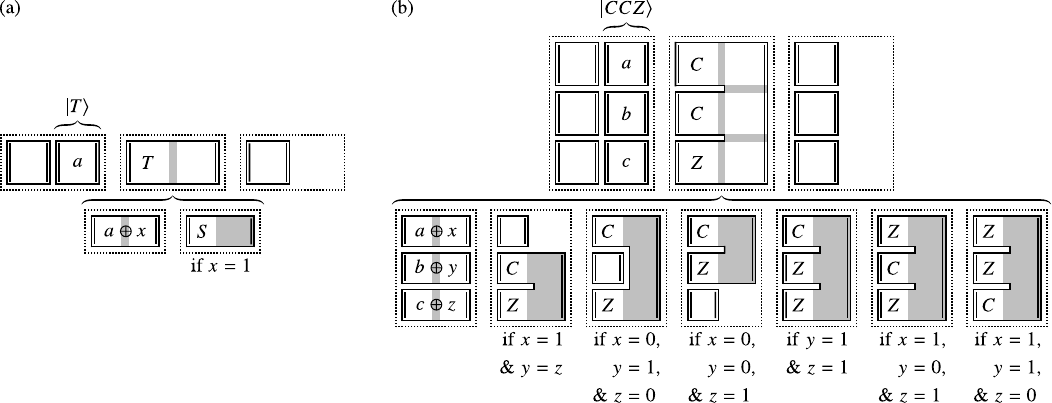}
\caption{Teleportation of non-Clifford gates (a) $T$ and (b) $CCZ$ by consuming the corresponding magic states $|T\rangle$ and $|CCZ\rangle$.
In both cases, the patch merging operation performs the logical $X$ correction on the magic state to stabilize the merged patch.
Magic states may also have a non-trivial Clifford frame that modifies the Clifford correction.
The trivial Clifford frame is $a = b = c = 0$.
\vspace*{-1.4mm}
}\label{fig_teleport}
\end{figure*}

Like Fig.\@ \ref{fig_clifford}, the Clifford gates in Fig.\@ \ref{fig_multi} can be described by the connectivity of timelike twist defects across the operation.
However, some twist defects are reflected either forwards or backwards in time, so the operation is no longer just a permutation of the initial defects.
These reflections also complicate the heuristic propagation of Pauli $Y$ strings attached to twist defects across the Clifford gate.
Valid Pauli strings can only pass through the interior of surface code patches, so strings between twist defects in two patches must be mediated by connections to intermediate defects shared by patches.
For example, an initial string connecting defects $1$ and $2$ in Fig.\@ \ref{fig_multi}c propagates to a final string connecting defects $1$ and $b$, then $b$ and $c$, and finally $c$ and $2$.
This corresponds to a logical $X$ string on the control patch accumulating logical $Z$ strings on the target patches after propagating through the $CZ^2$ gate.

\subsection{Non-Clifford gates}
\vspace*{-0.25cm}

Logical non-Clifford gates are usually implemented on the surface code by gate teleportation, as depicted in Fig.\@ \ref{fig_teleport} for the $T$ and $CCZ$ gates.
As suggested by a ZX calculus analysis \cite{zx_calculus}, these gates use patch merging operations to entangle the magic state with the target state.
Just as with the multi-qubit Clifford gates in Fig.\@ \ref{fig_multi}, the $CCZ$ gate in Fig.\@ \ref{fig_teleport}b is straightforward to adapt to separated control and target patches.
However, this separation now increases the cost of multiple $CZ$ gates during the $CCZ$ gate, and it may be more efficient to move control and target patches together before performing a $CCZ$ gate.
A spatially local $CCZ$ gate also avoids moving the patches storing $|CCZ\rangle$ to the control and target patches and may allow a $|CCZ\rangle$ to be consumed closer to where it was prepared.

It is possible to increase the speed of the $CCZ$ gate in Fig.\@ \ref{fig_teleport}b by adding another ancilla region on the left side of the computational region.
This increases the space cost from $6 d^2$ to $9 d^2$ data qubits.
The larger ancilla space allows for all $CZ$ corrections to be performed simultaneously, although the schedule of entire domain walls needs adjustment.
The expected time cost is reduced from $3.125 d$ to $2.75 d$ error correction cycles, and the maximum time cost is reduced from $5 d$ to $3 d$.
This results in a marginal scenario where the spacetime cost either increases from $18.75 d^3$ to $24.75 d^3$ qubitcycles or decreases from $30 d^3$ to $27 d^3$ depending on whether the expected or maximum time is applicable.
The expected time is relevant for adaptive scheduling that can benefit from the early termination of an operation, while the maximum time is relevant for rigid scheduling that must wait for the completion of many stochastic operations.

The gates in Fig.\@ \ref{fig_teleport} are also compatible with Clifford frames \cite{clifford_frame} carried by the magic states.
A Clifford frame describes a delayed Clifford operation that can be combined with a future Clifford operation if they are more efficient to implement together.
While not strictly necessary, this feature enables a free Clifford twirling operation to be applied to magic states prior to gate teleportation, which converts coherent errors into stochastic errors \cite{distillation1}.
I discuss Clifford twirling of magic states in Sec.\@ \ref{analysis_section}.

The combination of the Clifford and non-Clifford gates presented in this section form a universal gate set on a linear array of surface code tiles that are tightly packed along their rough boundaries.
All of the gates in this set need ancilla patches adjacent to one smooth boundary of the surface code tiles that they act upon.
The non-Clifford gates also consume magic states, so operations to prepare magic states are needed to enable this gate set.
The rest of this paper is focused on the preparation of $|T\rangle$ and $|CCZ\rangle$.
I summarize these operations and their costs in Table \ref{table_cost}.

\begin{table*}
\caption{\label{table_cost} Universal gate set on a linear array of distance-$d$ surface code tiles packed tightly along their rough boundary with space for an ancilla tile next to one smooth boundary. $L$ is the number of ancilla tiles in a path connecting the tiles in a multi-qubit gate.}
\begin{ruledtabular}
\begin{tabular}{lllll}
Gate & Space cost (data qubits) & Time cost (error correction cycles) & Resource cost & Implementation \\
\hline
$S$ & $2d^2$ & $2d$ & -- & Fig.\@ \ref{fig_clifford}a \\
$H$ & $2d^2$ & $3d$ & -- & Fig.\@ \ref{fig_clifford}b \\
$SH$ & $2d^2$ & $3d$ & -- & Fig.\@ \ref{fig_clifford}c \\
$HS$ & $2d^2$ & $3d$ & -- & Fig.\@ \ref{fig_clifford}d \\
$SHS$ & $2d^2$ & $4d$ & -- & Fig.\@ \ref{fig_clifford}e \\
$CZ^n$ (one-sided) & $(L+n+1)d^2$ & $2d$ & -- & Fig.\@ \ref{fig_multi}c \\
$CZ^n$ (two-sided) & $(L+n+1)d^2$ & $3d$ & -- & Fig.\@ \ref{fig_multi}d \\
\hline
$T$ & $2d^2$ & $2d \pm 1d \le 3d$ & $|T\rangle$ & Fig.\@ \ref{fig_teleport}a \\
$CCZ$ (narrow) & $(L+3)d^2$ & $3.125d \pm 1.95d \le 5 d$ & $|CCZ\rangle$ & Fig.\@ \ref{fig_teleport}b \\
$CCZ$ (wide) & $(2L+3)d^2$ & $2.75d \pm 0.661d \le 3d$ & $|CCZ\rangle$ & -- \\
$|T\rangle$ preparation & $3d^2$ & $15d$ & -- & Fig.\@ \ref{fig_T} \\
$|CCZ\rangle$ preparation & $6d^2$ & $10.5d$ & -- & Fig.\@ \ref{fig_CCZ}
\end{tabular}
\end{ruledtabular}
\end{table*}

\subsection{Error model \label{error_subsection}}
\vspace*{-0.25cm}

For expediency, the error analysis performed in this paper uses a simple error model for surface code operations.
Each time interval of $d$ error correction cycles on each active distance-$d$ surface code tile is considered as a possible location for an independent error at the end of the interval.
A more precise error model would involve a geometric analysis of the operation accounting for the number of endpoints for each type of logical error string to determine its error probability.
An even more precise but computationally expensive model would perform a stabilizer simulation on top of a vector state simulation of the logical subspace to simulate physical errors at the circuit level.
The cost of these direct simulations may be reduced by rare event sampling \cite{rare_events}.
I defer this more precise error modeling to future work.

Within each logical error location, I consider spacelike $X$ and $Z$ errors and timelike measurement errors as independent events with probability
\begin{equation}
 p_L(d) = C(d) p^{ \lfloor (d+1)/2 \rfloor }
\end{equation}
 for a code distance $d$ and a physical error probability $p$.
The factor $p^{ \lfloor (d+1)/2 \rfloor }$ is the probability of a pattern of $\lfloor (d+1)/2 \rfloor$ physical errors that the code cannot distinguish from another pattern of $\lfloor (d-1)/2 \rfloor$ physical errors.
The combination of these indistinguishable patterns is a logical error string.
The factor $C(d)$ is the number of patterns within the error location that cause these logical errors.
I use the specific empirical model form
\begin{equation} \label{specific_error_model}
 p_L(d) = p_{L @ T}^{d \bmod 2} (p/p_T)^{ \lfloor (d+1)/2 \rfloor }
\end{equation}
 for an error threshold $p_T$ and a logical error rate at threshold $p_{L@T}^b$ that depends on the parity of the code distance $b$.
I fit values $p_T=0.007$, $p_{L@T}^0 = 0.09$, and $p_{L@T}^1 = 0.3$ to the small logical error rates from numerical simulations of the rotated surface code with depolarizing noise and a minimum-weight perfect matching decoder \cite{error_model}.

\section{Design\label{design_section}}
\vspace*{-0.25cm}

I now consider the design of magic state distillation operations within the surface code lattice surgery framework.
I first consider the preparation of $|T\rangle$ on a distance-$d$ surface code tile by distilling 15 input $|T\rangle$ on distance-$\lfloor d/3 \rfloor$ surface code tiles.
To simplify the consideration of multi-level distillation in this section, I assume that $d$ is power of 3 times a base code distance in which physical $|T\rangle$ are injected.
I begin with a brief review of existing distillation designs, then consider the minimum requirements of fault tolerance, and finally design a detailed spacetime schedule of surface code operations.
I then adapt this design to the preparation of $|CCZ\rangle$ on three distance-$d$ surface code tiles by distilling 8 input $|T\rangle$ on distance-$\lfloor d/2 \rfloor$ surface code tiles.
In addition to the previous assumptions, I also assume that $d$ is divisible by 2 for $|CCZ\rangle$.

To bound the use of space in multi-level 15-to-1 distillation of $|T\rangle$, I consider a generic recursive structure for all designs.
The only requirement is an $L d$-by-$W d$ computational region for integers $L$ and $W$ in which one level of distillation can be implemented alongside the packed storage of its 15 $|T\rangle$ inputs.
To distill $|T\rangle$ up to distance $d$, I first partition the computational region into a uniform 3-by-3 grid and simultaneously distill 9 $|T\rangle$ up to distance $\lfloor d/3 \rfloor$.
I then pack the 9 $|T\rangle$ into 3 of the subregions, and simultaneously distill 6 more $|T\rangle$ up to distance $\lfloor d/3 \rfloor$ in the empty subregions.
Finally, I move the remaining 6 $|T\rangle$ into their packed location and perform the next level of distillation up to distance $d$.
Ignoring failed distillation steps until the next section, the time cost of $m$ levels of distillation is
\begin{equation} \label{multi_cost}
 t_{\mathrm{multi}}(m) =  \sum_{r=0}^{m-1} (2/3)^r (t_{\mathrm{distill}} + t_{\mathrm{pack}}) - t_{\mathrm{pack}},
\end{equation}
 where $t_{\mathrm{distill}}$ and $t_{\mathrm{pack}}$ are the time costs of the distillation and packing steps at distance $d$.
Each level of recursion introduces twice as many batches of distillation steps, but they are three times faster because of their smaller code distance.
Formally, the cost of infinite levels of recursion is finite,
\begin{equation} \label{multi_cost_bound}
 t_{\mathrm{multi}}(m) \le  t_{\mathrm{multi}}(\infty) = 3 t_{\mathrm{distill}} + 2 t_{\mathrm{pack}},
\end{equation}
 since the infinite geometric series converges.
This simple accounting assumes that physical $|T\rangle$ injection at the base level of recursion has the same time cost as consuming stored $|T\rangle$ inputs.

The original design for 15-to-1 distillation on the surface code \cite{top_down1} uses the transversal $T$ gate of the Reed-Muller code to teleport a distilled $|T\rangle$ into the surface code.
It is implemented with defect-based qubits and braiding-based $CX^n$, but the implementation is straightforward to adapt to tile-based qubits and the gate set in Table \ref{table_cost}.
Because $CX^n$ gates spread errors, all operations must be performed on a distance-$d$ surface code for fault tolerance.
The distillation circuit and $|T\rangle$ storage space fit in a 2-by-17 array of surface code tiles, with separate rows for persistent qubits and ancilla space.
The circuit can be modified to use 1-sided $CZ^n$ gates instead of $CX^n$ gates, with an initial preparation of $|+\rangle$ and a Bell state and two extra time intervals for $H$ gates.
The 15 distance-$\lfloor d/3 \rfloor$ $|T\rangle$ inputs can be packed into the free column, and I assume that they can be moved into place for teleportation without slowing down the circuit.
The space cost is $34 d^2$, and the most likely time cost is $20 d$ for Bell preparation, 5 $CZ^n$ gates, 2 $H$ gates, and concurrent $T$ teleportation.
Each packing of $|T\rangle$ outputs can be performed in one data movement step for a total time cost of $2d$.
Thus the spacetime cost of $|T\rangle$ distillation is $680 d^3$ qubitcycles for one level and $2176 d^3$ qubitcycles for infinite levels.

Litinski's design for 15-to-1 distillation on the surface code \cite{distillation4} avoids the Reed-Muller encoding step and reduces the number of persistent qubits from 16 to 5.
His distillation circuit is a sequence of 15 multiqubit $\pi/8$ $Z$ rotations between $|+\rangle$ preparation and $X$ measurement that increases concurrency and fault tolerance.
A persistent $d$-by-$d$ surface code tile is needed to store the output $|T\rangle$, but anisotropic $d$-by-$\lfloor d/3 \rfloor$ surface code tiles can be used for the four other persistent qubits.
The circuit detects up to two $Z$ errors on these four tiles, so they only need a distance of $\lfloor d/3 \rfloor$ with respect to $Z$ errors.
The computational region for one level of multi-level 15-to-1 distillation fits in a $3d$-by-$5d$ grid of data qubits.
It has a central $3d$-by-$7 \lfloor d/3 \rfloor$ grid for the row of persistent qubits between two ancilla rows for Pauli measurements of the same width to preserve the code distance of the output tile.
It has an additional four ancilla patches on the sides to teleport $T$ gates and perform concurrent Clifford corrections, with enough extra space to store the 15 $|T\rangle$ inputs.
This layout enables the 15 $\pi/8$ $Z$ rotations to be applied with a total time cost of $8 \lfloor d/3 \rfloor$, although this requires internal timelike twist defects to implement logical $Y$ measurements.
Again, each packing of $|T\rangle$ outputs can be performed in one data movement step for a total time cost of $2d$.
Litinski's design thus reduces the spacetime cost of $|T\rangle$ distillation to $40 d^3$ qubitcycles for one level and $180 d^3$ qubitcycles for infinite levels.

My goal of a new design for 15-to-1 distillation on the surface code is to use the Reed-Muller code to detect errors in both the $|T\rangle$ inputs and all distance-$\lfloor d/3 \rfloor$ surface code operations.
The 15 $|T\rangle$ inputs can be directly encoded in the Reed-Muller code by measuring its $Z$ stabilizer generators and adjusting the random Pauli frame.
The encoded $|T\rangle$ state can then be teleported into a distance-$d$ surface code tile initialized to $|+\rangle$ with a logical $Z$ pair measurement followed by $X$ measurements on all tiles in the Reed-Muller code.
The terminal $X$ measurements detect all problematic $Z$ errors, while redundancies in the $Z$ measurements can be added to detect problematic $X$ errors.
The two challenges of this design are identifying sufficient fault tolerance criteria for the $X$ errors and optimizing the schedule and layout to minimize its spacetime cost.

An important design freedom in the scheduling of stabilizer measurements is the ability to measure stabilizers before preparing all encoded qubits and after measuring some encoded qubits.
In particular, the distance-$d$ surface code tile containing the $|T\rangle$ output occupies a lot of space.
It is beneficial to measure as many qubits as possible in the Reed-Muller code before preparing the distance-$d$ tile to reduce the size of the computational region.
An example of this possibility is the fault tolerant conversion from the Reed-Muller code on 15 qubits to the Steane code on 7 qubits \cite{steane}.
Gate teleportation from the Reed-Muller code is a destructive process that ends with all of its qubits measured, and measuring them as soon as possible is likely to reduce spacetime costs.
This is a partially frustrated design optimization because fault tolerance requirements restrict when qubits can be measured, depending on which qubits have already been measured.

\subsection{Fault tolerance requirements}
\vspace*{-0.25cm}

To understand the fault tolerance requirements of $|T\rangle$ distillation, I consider a tesseract with vertices labeled by four bits, each corresponding to a dimension.
The $0000$ vertex is a distance-$d$ surface code tile, and the remaining 15 vertices are distance-$\lfloor d/3 \rfloor$ surface code tiles.
I use the vertex labels to label the logical operators of the tiles, and I use asterisks to denote products of operators such as
\begin{equation}
 X_{0**1} = \prod_{x,y \in \{0,1\} } X_{0xy1}.
\end{equation}
The logical Pauli operators of the distance-$d$ tile are $X_{0000}$ and $Z_{0000}$.
Next, I consider the 15 distance-$\lfloor d/3 \rfloor$ tiles to be encoded in the $[[15,1,3]]$ Reed-Muller code \cite{distillation1}.
A canonical choice of operators for this code is $X_{0000} X_{****}$ as logical $X$, $Z_{0000} Z_{****}$ as logical $Z$,
\begin{equation} \label{canonical_X}
 \{ X_{1***}, X_{*1**}, X_{**1*}, X_{***1} \}
\end{equation}
 as the generators of the $X$-type stabilizer subgroup, and
\begin{align} \label{canonical_Z}
 \{ & Z_{1***}, Z_{*1**}, Z_{**1*}, Z_{***1}, Z_{11**}, \notag \\
 & Z_{1*1*}, Z_{1**1}, Z_{*11*}, Z_{*1*1}, Z_{**11} \}
\end{align}
 as the generators of the $Z$-type stabilizer subgroup.
This tesseract encodes two distance-$d$ logical qubits, which I restrict to a Bell state by adding the pairs of logical operators $X_{****}$ and $Z_{****}$ to the stabilizer group.

I refer to the measurement outcomes of the stabilizer generators as the Pauli frame of the code.
Stabilizer-based error correction detects errors from changes in the Pauli frame, and the specific value of the Pauli frame is usually irrelevant.
The all-zero Pauli frame can be recovered by applying a Pauli destabilizer operator, but this operation can be tracked in software and deferred.
However, some implementations of non-Clifford gates do depend on specific values in part of the Pauli frame to preserve an expanded, non-Abelian stabilizer group \cite{distillation1}.
I consider the generic case of a state $|\Psi\rangle$ that is uniquely defined up to a constant by a stabilizer group $\mathcal{G}$, satisfying
\begin{equation}
 g |\Psi\rangle = |\Psi\rangle
\end{equation}
 for all $g \in \mathcal{G}$.
Then $|\Psi\rangle$ is also a right eigenstate of an operator $P$ if for every $g \in \mathcal{G}$ it satisfies
\begin{equation} \label{commute}
 g P = P g'
\end{equation}
 for some $g' \in \mathcal{G}$.
This follows from a test of the stability of $P |\Psi\rangle$ for every $g \in \mathcal{G}$,
\begin{equation} \label{stable_test}
 g (P |\Psi\rangle ) = P g'|\Psi\rangle = P|\Psi\rangle .
\end{equation}
If $P|\Psi\rangle$ is stabilized by all $g$, then it must be proportional to $|\Psi\rangle$.
Unless $|\Psi\rangle$ is in the null space of $P$, the corresponding eigenvalue can be set to one by redefining $P$ with a multiplicative prefactor.
After this rescaling, $P$ becomes a non-Abelian stabilizer of $|\Psi\rangle$ that requires the preservation of the all-zero Pauli frame to satisfy Eq.\@ (\ref{stable_test}).
These simple requirements are not strictly necessary, and $|\Psi\rangle$ only needs to be an eigenvector of $g$ and $g'$ in Eq.\@ (\ref{commute}) with the same eigenvalue.
As a result, only part of the all-zero Pauli frame needs to be preserved, depending on details of the commutation structure of $\mathcal{G}$ with $P$.

In the specific case here of a Bell state stabilized on a tesseract, its $Z$-type stabilizer subgroup can be expanded into a non-Abelian stabilizer subgroup defined by generators
\begin{equation} \label{non_abelian}
 \{T_{****}, S_{1***}, S_{*1**}, S_{**1*}, S_{***1} \},
\end{equation}
 which generate $Z_{****}$ and Eq.\@ (\ref{canonical_Z}).
These generators commute with elements of the $Z$-type stabilizer subgroup and produce terms from the non-Abelian subgroup when commuted with elements of the $X$-type stabilizer subgroup.
Representative, non-trivial instances of Eq.\@ (\ref{commute}) are
\begin{align} \label{T_stabilizer}
  X_{1***} S_{1***} &= S_{1***} (Z_{1***} X_{1***}), \notag \\
  X_{1***} S_{*1**} &= S_{*1**} (Z_{11**} X_{1***}), \notag \\
  X_{1***} T_{****} &= T_{****} (S_{1***}^\dag X_{1***}),
\end{align}
 with $S^\dag = S Z$.
The operators in Eq.\@ (\ref{non_abelian}) are also diagonal operators in the computational basis that trivially preserve the all-zero basis state, thus they are stabilizers.
The $T_{****}$ stabilizer in particular enables $T$ gate teleportation through the equivalence of applying $T^\dag$ to the $0000$ tile with applying $T$ to the 15 other tiles in the tesseract.

I assume that the stabilizer-defined state of the tesseract is implemented by preparing the $0000$ tile as $|+\rangle$ and the 15 other tiles as $|T\rangle = T |+\rangle$ and then measuring the $Z$-type stabilizer generators.
In the absence of errors, this results in an all-zero $X$ Pauli frame and a random $Z$ Pauli frame.
After correcting the $Z$ Pauli frame to all all-zero state, a logical $X$ measurement on the Reed-Muller code results in $X |T\rangle$ on the $0000$ tile after outcome $0$ or $X Z |T\rangle$ after outcome $1$.
The desired output $|T\rangle$ can then be recovered by a Pauli correction to the $0000$ tile.

The random $Z$ Pauli frame can be adjusted to an all-zero state by applying a Pauli correction constructed from a product of $X$-type destabilizer generators that each flip one bit.
However, the 15 $|T\rangle$ inputs correspond to $T$ applied before the Pauli frame correction, while the simple non-Abelian stabilizer structure only exists after the correction.
The $T$ gates can be effectively commuted to the end of the circuit, but their commutation with $X$ corrections on a tile transforms $T$ into $\sqrt{i} T^\dag$.
If a Clifford frame correction of $S^\dag X$ is used instead of $X$ on each corrected tile, then the $T^\dag$ gate on the $0000$ tile is restored.
Similarly, the Clifford frame correction on the 15 tiles can be negated to apply a $T$ gate on the $0000$ tile instead.

I also assume that the terminal logical $X$ measurement is implemented by independent $X$ measurements of each tile in the Reed-Muller code.
This implementation enables the measurement of all $X$-type stabilizer generators to be reconstructed in addition to the logical $X$ measurement.
These stabilizer measurements can detect up to two $Z$ errors as changes to the all-zero $X$ Pauli frame.
There is a formal correspondence between data and measurement errors when stabilizer measurements are reconstructed from single-qubit measurements, therefore $X$ measurement redundancy is not needed.
A Clifford frame correction applied immediately before an independent $X$ measurement can be switched from $S^\dag X$ to $S$ since $S^\dag X = -i XS$ and $-i X$ can be absorbed into the measurement.
Alternatively, a conditional $S$ gate followed by an $X$ measurement can be implemented as a conditional measurement of $X$ or $Y$ \cite{diagonal_s_gate}.
Without explicit $X$ corrections, the $S$ corrections can be interpreted as adapting the $T$ stabilizer structure in Eq.\@ (\ref{T_stabilizer}) to the random $Z$ Pauli frame rather than switching to the all-zero $Z$ Pauli frame.
Measurement errors or $X$ errors that cause errors in the $Z$ Pauli frame measurement lead to erroneous frame corrections that correspond to $S$ or $S^\dag$ errors.
Because
\begin{equation} \label{coherent_error}
 S = \sqrt{i/2} I + \sqrt{-i/2} Z,
 \end{equation}
 an $S$ or $S^\dag$ error is effectively a coherent superposition of no error and a $Z$ error.
Changes to the $X$ Pauli frame can detect this type of error on up to two tiles.

Errors in the $Z$ Pauli frame can be detected with redundant measurements.
These 11 bits of information can be protected by adding redundant check bits corresponding to measurements of other $Z$-type operators from the underlying group of operators.
The simplest form of redundancy is to measure each stabilizer multiple times, but it is more efficient to encode the stabilizers with a classical linear code \cite{temporal_encoding}.
For example, up to 3 errors can be detected in 11 bits by adding 4 redundant bits with the $[15,11,3]$ Hamming code.
These redundant bits can be formed by right-multiplying a vector of the 11 measurement outcome bits with an encoding matrix using Boolean linear algebra,
\begin{align} \label{hamming}
\left[
\begin{array}{ccccccccccc}
 1 & 1 & 1 & 0 & 0 & 0 & 1 & 1 & 1 & 0 & 1 \\
 1 & 0 & 0 & 1 & 1 & 0 & 1 & 1 & 0 & 1 & 1 \\
 0 & 1 & 0 & 1 & 0 & 1 & 1 & 0 & 1 & 1 & 1 \\
 0 & 0 & 1 & 0 & 1 & 1 & 0 & 1 & 1 & 1 & 1
\end{array}
\right]
\end{align}
These bits can be measured directly as the operator product of stabilizer generators corresponding to the summed bits.
The redundant generators are determined by the specific choice of $Z$-type stabilizer generators and order of bits in the Hamming code.
For example, the redundant stabilizers corresponding to $Z_{****}$ and Eq.\@ (\ref{canonical_Z}) are
\begin{equation}
 \{ Z_{*000} Z_{*111}, Z_{0*00} Z_{1*11}, Z_{00*0} Z_{11*1}, Z_{000*} Z_{111*} \}
\end{equation}
 if the bit order is selected by associating 1's and 0's in each column of Eq.\@ (\ref{hamming}) to summed and unsummed indices.
The redundant bits must be measured independently, but they can be further decomposed into measurements of other operators in the stabilizer group.

The most complicated fault tolerance requirements come from the detection of $X$ errors.
To achieve an overall error detection distance of 3, the distillation circuit must detect up to two $X$ errors in the absence of measurement errors and one $X$ error in the presence of one measurement error.
In a circuit containing only $Z$-type stabilizer measurements between $|T\rangle$ preparation and $X$ measurement, an $X$ error can be converted into measurement errors by commuting it to the beginning or end of the circuit.
The $X$ error is then fully absorbed into the $X$ measurement or converted into an $S^\dag$ error by the initial $|T\rangle$, which is stabilized by $\sqrt{-i} SX$.
This conversion into measurement errors depends on the order of stabilizer measurements.
Note that delayed measurements as in Fig.\@ \ref{fig_surgery}e can be used to avoid a strict time order of stabilizer measurements, and multiple stabilizer measurements can occur simultaneously.
Depending on the choice of commutation direction, each $X$ error location partitions the set of stabilizer measurements into correct and incorrect measurements.
Each pair of $X$ error locations induces a similar partition for each combination of commutation directions, with the additional behavior that a stabilizer measurement is correct if it is commuted with two $X$ errors.
A non-trivial error pattern is detectable if the measurement errors cause an inconsistency in the reconstruction of the $Z$ Pauli frame.
The choice of commutation direction determines if an error pattern is trivial, but it does not affect consistency because an $X$ error propagated through the entire circuit causes a consistent change to the Pauli frame.
An undetectable error that has a trivial pattern with no errors is irrelevant because any circuit-equivalent non-trivial pattern corresponds to either detectable $S^\dag$ errors or no output error.

The measurement-based $|T\rangle$ distillation circuits described here are fault tolerant if the stabilizer measurements satisfy the distinct criteria for two measurement errors, two $X$ errors, and one of each error.
Each error pattern corresponds to an overdetermined linear system in Boolean linear algebra that reconstructs the $Z$ Pauli frame, and an error is detectable if the system is not solvable.
An undetectable, non-trivial error pattern that is equivalent to a trivial pattern is irrelevant to the fault tolerance requirements, but it may spread into $S^\dag$ errors on up to two tiles.
Changes to the $X$ Pauli frame can detect these remaining errors in addition to all combinations of up to two errors that include $Z$ errors.
The next goal is to design a fault-tolerant implementation that minimizes the spacetime cost, which is distinct from minimizing the number or weight of stabilizer measurements.
The design space is complicated enough that global minimization of spacetime cost is beyond the scope of this paper.
Instead, I randomly sample from design space and select the fault-tolerant design that minimizes a simple cost proxy.

\subsection{Distillation of $|T\rangle$}
\vspace*{-0.25cm}

Because 9 $|T\rangle$ must be stored in distance-$\lfloor d/3 \rfloor$ surface code tiles in the computational region of 3 distance-$\lfloor d/3 \rfloor$ distillation circuits, the computational region must occupy at least 3 of these tiles.
For my design, I consider a minimal computational region containing a $d$-by-$3d$ grid of data qubits to prepare $|T\rangle$ on a distance-$d$ surface code tile.
Rather than waiting until all 15 $|T\rangle$ inputs are available, I consider three distinct stages of distillation.
In the first stage, some distillation steps are applied to the initial 9 $|T\rangle$ before the remaining 6 $|T\rangle$ are prepared.
In the second stage, all 15 $|T\rangle$ inputs are available to the distillation process.
In the third stage, the distance-$d$ output tile is prepared after measuring 6 distance-$\lfloor d/3 \rfloor$ input tiles.
Because the output tile occupies a third of the computational region, there is insufficient space for efficient stabilizer measurements while all 15 input tiles are active.

A design must choose which elements of the $Z$ stabilizer subgroup to measure and what order to measure them in.
While this group contains non-trivial elements with even weights between 4 and 12 and one weight-16 element, it can be generated from only weight-4 generators.
Fault tolerance considerations generally favor the use of generators with low weight to reduce the number of error locations that they are sensitive to.
If a high-weight stabilizer is decomposed into a product of weight-4 stabilizers, then its composite measurement still contains some useful information about the $Z$ Pauli frame if at least one measurement is correct.
The only benefit to direct measurement of a higher-weight stabilizer is that it can be faster for a specific spatial layout if the equivalent product of weight-4 stabilizers cannot be measured simultaneously.
A minimal distance-3 encoding against measurements errors such as the redundant bits defined by Eq.\@ (\ref{hamming}) has more limited design flexibility.
It may not be possible to construct a minimal set of weight-4 redundant stabilizers from weight-4 stabilizer generators.

There is not yet any unifying theory of fault-tolerant circuit design, and there are many strategies for achieving fault tolerance in specific codes and circuit implementations.
High-distance codes like the surface code use simple repetition in time to encode each stabilizer generator measurement independently with circuits that limit the spread of errors during measurement \cite{hook_errors}.
Low-distance codes introduce additional design elements such as flag qubits \cite{flag_qubits} to detect the spread of errors through a fault-tolerant circuit more explicitly.
Here, the fault tolerance requirements are specifically focused on detecting errors in $Z$-type stabilizer measurements.
Encoded stabilizer measurements \cite{temporal_encoding} protect against measurement errors up to a certain code distance, but protection again $X$ errors depends on the circuit design.
This design may be assisted by the ability of the underlying Reed-Muller code to detect $X$ errors with respect to a known $Z$ Pauli frame.
The full $Z$ Pauli frame can detect up to 6 $X$ errors, while a more limited $Z$ Pauli frame associated with the subgroup generated by
\begin{equation}
 \{ Z_{1***}, Z_{*1**}, Z_{**1*}, Z_{***1} \}
\end{equation}
 can detect up to 2 $X$ errors in the same way that Eq.\@ (\ref{canonical_X}) can detect up to 2 $Z$ errors.
Perhaps there is a unifying theory for the simultaneous detection of measurement and data errors that simplifies this design process, although here I proceed with design in the absence of such a theory.

I explore the design space with software to test that a sequence of stabilizer measurements is sufficient to satisfy the fault tolerance requirements.
Furthermore, I test that all measurements are necessary by testing fault tolerance after removing each measurement from the sequence.
This enables the identification of locally optimal implementations in design space that are necessary and sufficient in this way.
It is straightforward to construct locally optimal implementations by adding measurements to a sequence if it is not sufficient and removing measurements if they are not necessary.
However, this does not provide enough structure to search efficiently for a globally optimal implementation that minimizes the number of stabilizer measurements or some other cost metric.
Here, I restrict the design process to a simple implementation that is locally optimal and adapt it to the prescribed computational region and three-stage distillation process.

I begin the design process with a few simple designs and empirical observations on random designs.
Repeating a minimal sequence of stabilizer generator measurements three times is a sequence of 33 measurements that is necessary and sufficient.
However, random sequences of stabilizer measurements require much fewer than 33 measurements to be sufficient.
As the number of measurements decreases, random circuits restricted to weight-4 stabilizer measurements are more likely to be sufficient.
The smallest number of weight-4 measurements that I observed to be sufficient was 17, with a probability of $(3.8 \pm 0.1) \times 10^{-6}$ over random sequences of 17 weight-4 measurements.
This is close to the lower bound of 15 set by the Hamming code, which is well known to be optimal for detecting up to two measurement errors.
I tested some Hamming encodings with sequences of 15 measurements, and they are not sufficient because they do not reliably detect $X$ errors.

To construct random designs that are compatible with a three-stage distillation process, I further constrain the distribution of stabilizer measurement sequences.
I constrain the first 14 weight-4 measurements in the sequence to exclude the 0000 tile, and I constrain the last 3 weight-4 measurements to include the 0000 tile.
In this constrained distribution, the probability of a sufficient sequence of 17 measurements decreases to $(6.36 \pm 0.08) \times 10^{-7}$.
Rather than add further constraints to the distribution and reduce this probability even more, I continue the design process by filtering sufficient sequences with extra design criteria.
The first stage of distillation with 9 tiles available can accommodate up to 4 measurements from these sufficient sequences with a probability of $(6.5 \pm 0.3) \times 10^{-2}$.
For the remaining 10 measurements in the second stage of distillation, I examined their packing into simultaneous clusters of disjoint measurements.
From the several hundred samples that I examined, only one sequence was able to pack into 4 such clusters, and I use this sequence as the starting point for the remaining design process.

\begin{table}
\vspace{-2.2mm}
\caption{\label{table_measurements} Fault-tolerant stabilizer measurement sequence for $|T\rangle$ distillation.
Here, tesseract vertices are labeled by integers corresponding to their 4-bit binary representation: $0000 = 0$, $0001=1$, ... , $1111=15$.
Each independent stabilizer has a destabilizer to adjust its bit in the $Z$ Pauli frame to zero.}
\begin{ruledtabular}
\begin{tabular}{lll}
Stage & Stabilizer & Destabilizer \\
\hline
 1a & $Z_{3} Z_{7} Z_{8} Z_{12}$ & $X_{0} X_{2} X_{8} X_{10} X_{14} X_{15}$ \\
 1b & $Z_{1} Z_{6} Z_{8} Z_{15}$ & $X_{4} X_{8} X_{12} X_{14}$ \\
 1c & $Z_{1} Z_{3} Z_{5} Z_{7}$ & $X_{0} X_{1} X_{4} X_{10} X_{11} X_{15}$ \\
 1d & $Z_{3} Z_{4} Z_{8} Z_{15}$ & $X_{0} X_{4} X_{8} X_{12} X_{14} X_{15}$ \\
 2a & $Z_{2} Z_{6} Z_{8} Z_{12}$ & $X_{2} X_{10}$ \\
 2a & $Z_{3} Z_{7} Z_{9} Z_{13}$ & $X_{8} X_{9} X_{10} X_{12} X_{14} X_{15}$ \\
 2b & $Z_{1} Z_{7} Z_{11} Z_{13}$ & $X_{11}$ \\
 2b & $Z_{2} Z_{4} Z_{10} Z_{12}$ & $X_{10}$ \\
 2b & $Z_{3} Z_{5} Z_{8} Z_{14}$ & $X_{14}$ \\
 2c & $Z_{1} Z_{4} Z_{9} Z_{12}$ &  $X_{0} X_{8} X_{10} X_{12} X_{14} X_{15}$ \\
 2c & $Z_{2} Z_{7} Z_{11} Z_{14}$ & -- \\
 2c & $Z_{3} Z_{6} Z_{10} Z_{15}$ & -- \\
 2d& $Z_{8} Z_{9} Z_{14} Z_{15}$ & -- \\
 2d & $Z_{2} Z_{4} Z_{11} Z_{13}$ & -- \\
 3a & $Z_{0} Z_{6} Z_{9} Z_{15}$ & $X_{0}$ \\
 3b & $Z_{0} Z_{5} Z_{11} Z_{14}$ & -- \\
 3c & $Z_{0} Z_{2} Z_{8} Z_{10}$ & -- \\
\end{tabular}
\end{ruledtabular}
\end{table}

\begin{figure*}
\includegraphics{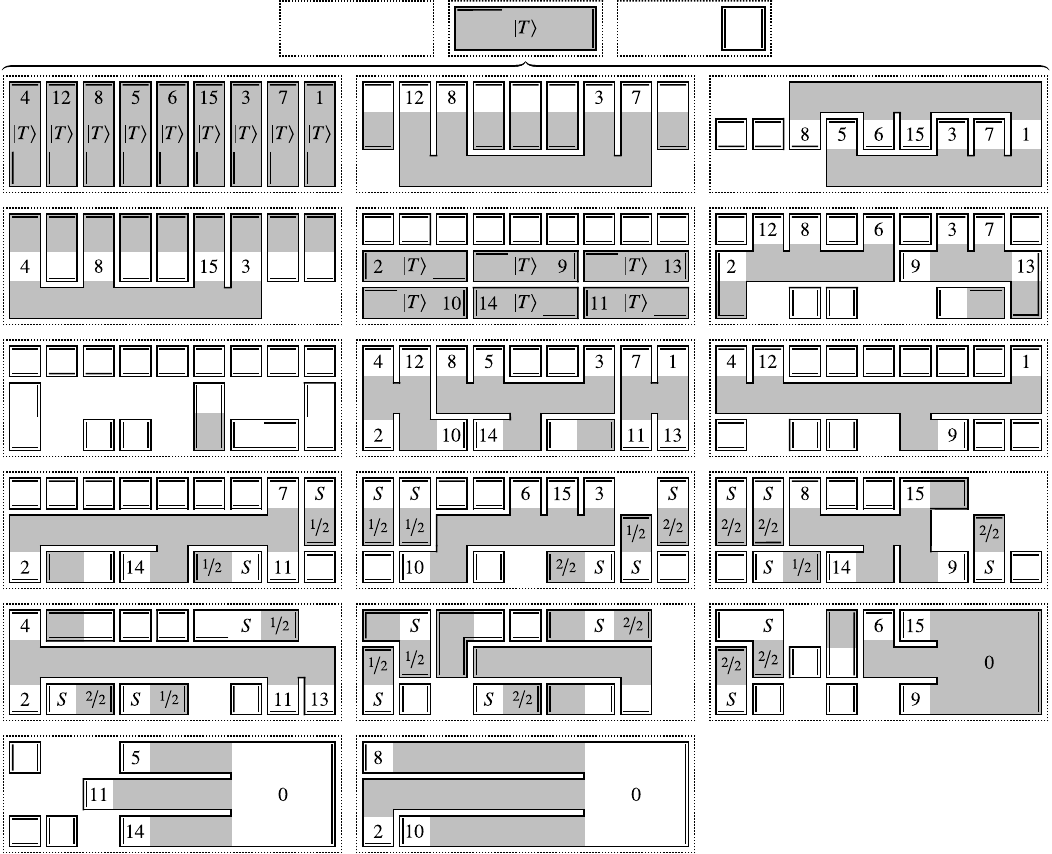}
\caption{Recursive multi-level distillation of $|T\rangle$. Each stabilizer measurement from Table \ref{table_measurements} is labeled with its constituent tesseract vertices for clarity.
The first 10 stabilizers measure the random $Z$ Pauli frame of the Reed-Muller code, and the first stabilizer that includes the output tile completes the measurement of the full $Z$ Pauli frame.
The remaining stabilizer measurements detect errors in the $Z$ Pauli frame.
The $S$ gates are used to adjust the transversal $T$ operation for the random $Z$ Pauli frame, and their implementation follows Fig.\@ \ref{fig_clifford}a and requires two time intervals.
The Clifford frame correction is determined by the destabilizers in Table \ref{table_measurements} and the substitution of $S$ for $X$ on the non-output tiles.
Depending on the specific value of the $Z$ Pauli frame and its correction, some of the $S$ gates depicted must be replaced by idle operations.
All non-output tiles are terminated by logical $X$ measurements.
Two bits of the $X$ Pauli frame are measured before the final time interval, and the remaining two bits are measured after it.
If any errors are detected by inconsistent stabilizer measurements, then this sub-tree of the $|T\rangle$ distillation process tree must start again from the beginning.
\vspace{-1.4mm}
}\label{fig_T}
\end{figure*}

In principle, the design of an efficient spatial layout of a distillation circuit is a discrete optimization problem.
For expediency, I perform a heuristic partial optimization here to achieve a good implementation that is unlikely to be optimal.
The main design frustration is the balance between data movement and simultaneous measurements.
The measurement sequence allows for 4 simultaneous clusters of measurements, with 2 clusters of 2 and 2 clusters of 3.
However, these clustered measurements cannot all be implemented with the same spatial layout of qubits, and extra data movement is required for full utilization of this efficiency.
A simple, static layout that enables stabilizer measurements is a row of ancilla tiles between two rows of active tiles.
From sampling random configurations in this layout, a static arrangement of tiles can achieve only 6 rounds of measurements in the second stage rather than 4.
Static arrangements that minimize data movement during the first stage can achieve only 7 rounds of measurements.
In this packed layout, extra data movement to improve concurrency does not reduce the time cost, and my design target is 7 rounds of measurement in the second stage.
With the remaining design freedom, I require that the first 2 measurements of the second stage be concurrent with data movement and that the last 3 measurements in the third stage be performed without reordering.
Only 8 configurations satisfy all of these constraints, and I chose one that made it easier to move data for the third stage.

After 10 stabilizers have been measured, there is enough information to set the $Z$ Pauli frame within the Reed-Muller code.
There is enough design flexibility in the destabilizer generators to omit 5 tiles from the frame correction operation, with some limitations to this choice imposed by solvability of the resulting Boolean linear system.
I choose to omit frame corrections on the vertices with integer labels 3, 5, 6, 7, and 13 to reduce congestion in scheduling the remaining corrections.
The resulting destabilizer generators are listed in Table \ref{table_measurements}.
The $S$ corrections within the Reed-Muller code commute with the $Z$-type stabilizer measurements and can be performed at any time after the $Z$ Pauli frame has been measured.
Since it has no initial $T$ gate or terminal $X$ measurement, the $Z$ Pauli frame is adjusted by $X$ rather than $S$ on the output tile.

Fig.\@ \ref{fig_T} is the end result of this design process.
Overall, I was able to schedule at least one stabilizer measurement in all but two time intervals of the distillation operation.
I combined data movement and patch deformation with stabilizer measurements and $S$ gates in a few steps to maximize concurrency.
Stages 1b and 1c from Table \ref{table_measurements} are measured simultaneously even though they are not disjoint because the first stage has two ancilla rows available in one time interval.
Because many qubits can be measured during the second stage of distillation, there is more ancilla space for data movement in the third stage.
However, there is the extra requirement of scheduling the $S$ corrections to adjust for the random $Z$ Pauli frame.
I was able to pack the $S$ gates such that they do not slow down the distillation operation, even though they are half as fast as the stabilizer measurement operations.
The code and data used in this design process are available on Zenodo \cite{zenodo}.

Using the same basic accounting as with previous distillation designs, one stage of this distillation process has a space cost of $3d^2$ data qubits and a time cost of $5 d$ error correction cycles.
The $|T\rangle$ packing cost for multi-level distillation in Eq.\@ (\ref{multi_cost}) is zero because it is integrated into the distillation process.
The spacetime cost of $|T\rangle$ distillation is $15d^3$ qubitcycles for one level and $45d^3$ qubitcycles for infinite levels.
Compared to Litinski's multi-level 15-to-1 design \cite{distillation4}, this design reduces the space cost by a factor of 5 and the spacetime cost by a factor of 4.

\subsection{Distillation of $|CCZ\rangle$}
\vspace*{-0.25cm}

Because it is based on a distance-2 code rather than a distance-3 code, it is much easier to design a distillation process for $|CCZ\rangle$.
Again, it can be defined by the stabilizer structure of an entangled state between input and output tiles.
The 8 input $|T\rangle$ are on distance-$\lfloor d/2 \rfloor$ tiles encoded in a [[8,3,2]] code, which I label with either 3 bits corresponding to vertices of a cube or the integer that they are the binary representation of.
The output $|CCZ\rangle$ is on 3 distance-$d$ tiles that I label with integers 8, 9, and 10.
Using the same product notation as before, the stabilizer group of the [[8,3,2]] code is generated by
\begin{equation}
 \{ X_{***}, Z_{***}, Z_{0**}, Z_{*0*}, Z_{**0} \},
\end{equation}
 and the three Bell states between input and output tiles are further stabilized by
\begin{align}
 \{  & X_{0**} X_8, X_{*0*} X_9, X_{**0} X_{10}, \notag \\
     & {-Z_{*00} Z_8}, -Z_{0*0} Z_9, -Z_{00*} Z_{10} \}.
\end{align}
Again, I consider a circuit that prepares the 8 input tiles as $|T\rangle$ and the 3 output tiles as $|+\rangle$, measures a sequence of $Z$ stabilizers, and then performs terminal $X$ measurements on all input tiles.
As before, this results in an all-zero $X$ Pauli frame in the absence of errors and a random $Z$ Pauli frame.
This circuit can detect a $Z$ error by reconstructing $X$-type stabilizer measurements, and redundant $Z$-type stabilizer measurements are needed for fault tolerance.

\begin{table}
\vspace{-2.2mm}
\caption{\label{table_measurements2} Fault-tolerant stabilizer measurements for $|CCZ\rangle$ distillation.
Each independent stabilizer has a destabilizer to adjust its bit in the $Z$ Pauli frame to zero.}
\begin{ruledtabular}
\begin{tabular}{lll}
Stabilizer & Destabilizer \\
\hline
 $-Z_{0} Z_{4} Z_{8}$ & $X_{0} X_{3} X_{7} X_{9}$ \\
 $-Z_{1} Z_{5} Z_{8}$  & $X_{5} X_{10}$ \\
 $-Z_{2} Z_{6} Z_{8}$  & $X_{0} X_{3} X_{5} X_{8} X_{9} X_{10}$ \\
 $-Z_{3} Z_{7} Z_{8}$  & $X_{7}$ \\
 $-Z_{0} Z_{2} Z_{9}$  & $X_{3} X_{7} X_{9}$ \\
 $-Z_{1} Z_{3} Z_{9}$  & $X_{3} X_{7}$ \\
 $-Z_{4} Z_{5} Z_{10}$  & $X_{10}$ \\
 $-Z_{6} Z_{7} Z_{10}$  & -- \\
\end{tabular}
\end{ruledtabular}
\end{table}

Again, the distillation process can be understood as expanding the $Z$-type stabilizer subgroup into a non-Abelian stabilizer subgroup with a dependence on the $Z$ Pauli frame.
For the all-zero $Z$ Pauli frame, the expanded subgroup is generated by
\begin{align}
 \{ & T_{***} C_{8} C_{9} Z_{10}, -S_{0**} C_{9} Z_{10}, & \notag \\ 
 &  {-S_{*0*} C_{8} Z_{10}}, -S_{**0} C_{8} Z_{9} \},
\end{align}
 where $C_a O$ denotes that the operator $O$ is controlled on the value of the qubit with the label $a$.
The representative, non-trivial instances of Eq.\@ (\ref{commute}) are
\begin{align}
 X_{*0*} X_9 (-S_{0**} C_{9} Z_{10}) &= (-S_{0**} C_{9} Z_{10}) ( -Z_{10} Z_{00*}) X_{*0*} X_9, \notag \\
 X_{0**} X_8 (-S_{0**} C_{9} Z_{10}) &= (-S_{0**} C_{9} Z_{10}) ( Z_{0**}) X_{0**} X_8, \notag \\
 X_{***} (-S_{0**} C_{9} Z_{10}) &= (-S_{0**} C_{9} Z_{10}) ( Z_{0**}) X_{***}, \notag \\
 X_{0**} X_8 T_{***} C_{8} C_{9} Z_{10} &= T_{***} C_{8} C_{9} Z_{10} ( - S^\dag_{0**} C_{9} Z_{10}) X_{0**} X_8, \notag \\
 X_{***} T_{***} C_{8} C_{9} Z_{10} &= T_{***} C_{8} C_{9} Z_{10} (S^\dag_{***}) X_{***}.
\end{align}
Again, the effects of a non-zero $Z$ Pauli frame on the stabilizer structure can be corrected by applying $S$ corrections to the input tiles and $X$ corrections to the output tiles.
In the all-zero $Z$ Pauli frame, a $T$ gate applied to each input tiles is equivalent to $CCZ$ applied to the output tiles, which prepares the $|CCZ\rangle$ output.
The $|CCZ\rangle$ output is then isolated up to a known Pauli correction by the $X$ measurements on the input tiles.

Fault tolerance requirements for the $Z$-type stabilizer measurements are simpler for distance-2 codes.
A parity check of the $Z$ Pauli frame is sufficient to detect one measurement error, which necessitates at least 8 stabilizer measurements in total.
One $X$ error is equivalent to at most one measurement error by commutation to the beginning or end of the circuit if each distance-$\lfloor d/2 \rfloor$ tile participates in at most three stabilizer measurements.
These requirements are satisfied by the stabilizers in Table \ref{table_measurements2} for any measurement order, using only weight-3 stabilizers that each contain one distance-$d$ tile.

\begin{figure}
\includegraphics{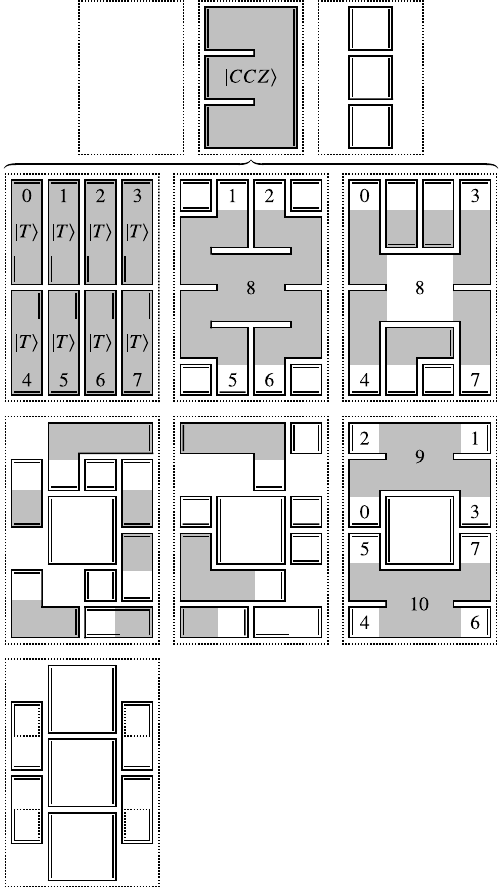}
\caption{Terminal distillation of $|CCZ\rangle$. Each stabilizer measurement from Table \ref{table_measurements2} is labeled for clarity.
All non-output tiles are terminated by logical $X$ measurements.
The random $Z$ Pauli frame is measured before the final time interval, and the zero $X$ Pauli frame is measured after it.
The Clifford frame correction is determined by the destabilizers in Table \ref{table_measurements2} and the substitution of $S$ for $X$ on the non-output tiles.
Depending on the required frame correction, some of the partial domain boundaries that implement $S$ gates in the final time interval must be omitted.
If any errors are detected by inconsistent stabilizer measurements, then this distillation process must start again from the beginning.
}\label{fig_CCZ}
\end{figure}

The layout and schedule of $|CCZ\rangle$ distillation in Fig.\@ \ref{fig_CCZ} is more straightforward than $|T\rangle$ distillation.
I prepare all 8 $|T\rangle$ inputs simultaneously in a $2d$-by-$3d$ spatial layout, which is also sufficient space to implement the distillation process.
Like in $|T\rangle$ distillation, I schedule as many measurements as possible before preparing all output tiles.
In this case, half of the distillation process can be completed with a single output tile, and the other half requires the remaining two output tiles.
Because both smooth boundaries of the output tiles are accessible, this design is able to measure all stabilizers concurrently in pairs.
Again, the destabilizer generators are chosen to select the most convenient locations for $S$ gate corrections to the distance-$\lfloor d/2 \rfloor$ tile.
Unlike in $|T\rangle$ distillation, it is more efficient to use a customized $S$ gate implementation rather than Fig.\@ \ref{fig_clifford}a.

Using the same basic accounting as with previous distillation designs, the $|CCZ\rangle$ distillation process has a space cost of $6d^2$ data qubits and a time cost of $3 d$ error correction cycles.
The $|T\rangle$ preparation time is nominally $15 \lfloor d/2 \rfloor$ cycles for infinite levels of $|T\rangle$ distillation.
The overall spacetime cost of $|CCZ\rangle$ distillation is then $63d^3$ qubitcycles.
A comparable $|CCZ\rangle$ distillation design with one level of $|T\rangle$ distillation \cite{distillation3} has a space cost of $72 d^2$ qubits and a total time cost of $5.5d$ cycles, which corresponds to a spacetime cost of $396 d^3$ qubitcycles.

\section{Analysis\label{analysis_section}}
\vspace*{-0.25cm}

A precise analysis of the error and cost of $|T\rangle$ and $|CCZ\rangle$ distillation requires statistics from direct numerical simulations, which is beyond the scope of this paper.
Instead, I present a simple analysis that is intended to provide a foundation of understanding from which to interpret more complicated and precise results.

An important assumption of this analysis is a Pauli error model as discussed in Sec.\@ \ref{error_subsection}.
While high-distance codes are expected to decohere noise strongly by projecting Pauli noise components into different Pauli frames, Pauli error assumptions are less valid for low-distance codes and physical states.
Such concerns are most relevant at the lower levels of magic state distillation that work with low-distance codes and injected physical magic states.
As originally noted by Bravyi and Kitaev \cite{distillation1} and emphasized in recent work \cite{clifford_twirling}, magic states can be twirled by their Clifford stabilizers to decohere their errors into stochastic Pauli errors.
Clifford twirling operations can thus enhance the validity of Pauli error models.
Twirling can be integrated into gate teleportation with no additional cost by applying transversal Pauli operations to magic state inputs to randomize the Clifford frame as shown in Fig.\@ \ref{fig_teleport}.
In effect, the Pauli component of a Clifford stabilizer is applied explicitly while the Clifford component is applied implicitly as the conditional Clifford correction of the teleported gate.
Similarly, the distillation operations in Sec.\@ \ref{design_section} are designed for $|T\rangle$ inputs, but they can be adapted to a randomized mix of $|T\rangle$ and $X |T\rangle$ inputs by adjusting the Clifford frame corrections.

In Fig.\@ \ref{fig_T}, the errors in a $|T\rangle$ output come from errors in the $|T\rangle$ inputs and distance-$\lfloor d/3 \rfloor$ surface code operations.
If I consider the $|T\rangle$ error probability as a function of the output code distance, then a simple approximate model of the error propagation is
\begin{equation} \label{propagate_error}
 p_M(d) = 35 [ p_M(\lfloor d/3 \rfloor) + 67.5 p_L(\lfloor d/3 \rfloor) ]^3,
\end{equation}
 using the simple model of surface code errors from Eq.\@ (\ref{specific_error_model}).
In the standard analysis of 15-to-1 $|T\rangle$ distillation \cite{distillation1}, there are 35 undetectable configurations of input errors.
Here, each error in such a configuration can come from either an input error or a logical error on the tile during the distillation process.
The nominal duration of the distillation process is 15 time intervals for no recursion and 45 time intervals for infinite recursion.
A $Z$ error is equivalent to an input error, while an $X$ error is only converted to an input error half of the time, depending on which Pauli component of Eq.\@ (\ref{coherent_error}) is measured.
Eq.\@ (\ref{propagate_error}) uses a simple overestimate in the number of logical error locations, $67.5 = 45 \times 1.5$.
It is an overestimate because some input tiles are not active for the entire duration, and many contributing patterns of $X$ errors are detected by inconsistencies in the redundant stabilizer measurements.
Accounting for these effects, Eq.\@ (\ref{propagate_error}) will become a more complicated cubic polynomial in $p_M$ and $p_L$ to leading order.
Further adjusting for the varying duration of the second recursion in Fig.\@ \ref{fig_T} will cause the coefficients of the polynomial to depend on $d$.
Eq.\@ (\ref{propagate_error}) does not include stabilizer measurement errors, which contribute a separate, smaller leading-order error term.

A distillation process described by Eq.\@ (\ref{propagate_error}) has an error threshold, above which the $|T\rangle$ output has a higher error probability than the $|T\rangle$ inputs.
The threshold $p_0$ is defined by the smallest positive real solution to
\begin{equation}
 p_0 = 35 [ p_0 + 67.5 p_L(d_0) ]^3,
\end{equation}
 for a surface code distance $d_0$ at the base level of distillation.
Alternatively, for any input error probability below the ideal threshold value, $p_0 < 0.169$, there is a constraint on the base code distance,
\begin{equation} \label{min_distance}
 d_0 > 2 \frac{\log\left(0.0151 p_0^{1/3} - 0.0494 p_0\right)}{\log(143 p)} - 1
\end{equation}
 for odd values of $d_0$, that must be satisfied for distillation to be effective.
If the base level is effective, then all higher levels of the distillation process with smaller $p_0$ and larger $d_0$ will also be effective.
This constraint will limit the effectiveness of distillation for physical error rates near the error threshold of the surface code and weaken as error rates are lowered well below threshold.
For example, $d_0 = 5$ satisfies Eq.\@ (\ref{min_distance}) for $p_0 = p = 7 \times 10^{-4}$.

\begin{figure}
\includegraphics{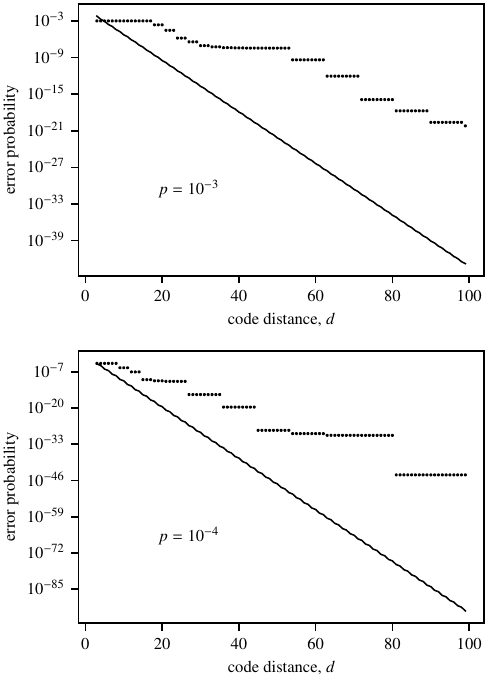}
\caption{Comparison between the error probability $p_M(d)$ of recursively distilled $|T\rangle$ (points) and the error rate $p_L(d)$ of the underlying surface code (solid line) for two physical error rates
 below the error threshold of the surface code but above the error threshold for compatible recursive distillation.
}\label{fig_error}
\end{figure}

Even when the multi-level distillation process reduces $p_M(d)$ for increasing $d$, it can still increase relative to the base $p_L(d)$ of the surface code.
I isolate an amplification $\Omega(d)$ relative to the underlying surface code error rate,
\begin{equation} \label{asymptotic_form}
 p_M(d) \sim \Omega(d) p_L(d),
\end{equation}
 which simplifies Eq.\@ (\ref{propagate_error}) to
\begin{equation} \label{amplification}
  \Omega(3d) = 450 p [\Omega(d)+ 67.5]^3
\end{equation}
 when $d$ is an odd multiple of 3.
In the asymptotic regime of $p \ll 1$, a constant function,
\begin{equation}
 \Omega(d) \sim (1.4 \times 10^{8}) p,
\end{equation}
 satisfies Eq.\@ (\ref{amplification}).
This is part of a range of $p$ values in which $p_M(d)$ remains proportional to $p_L(d)$ in the asymptotic regime of $d \gg 1$.
A constant $\Omega(d)$ reduces Eq.\@ (\ref{amplification}) to a cubic polynomial equation, which has a positive real solution for $p \le 7.2 \times 10^{-8}$.
Otherwise, $\Omega(d)$ eventually overwhelms 67.5 in Eq.\@ (\ref{amplification}) for $d \gg 1$, consistent with a growing lower bound,
\begin{equation}\label{amp_bound}
 \Omega(d) \ge \frac{\left[\sqrt{450 p} \Omega(d')\right]^{d/d'}}{\sqrt{450 p}},
\end{equation}
 for $d \ge d'$.
In this intermediate regime of
\begin{equation}
 7.2 \times 10^{-8} < p < 0.007,
\end{equation}
 the error probability of $|T\rangle$ is reduced by recursive distillation while still growing without bound relative to the logical error rate of the underlying surface code.

I evaluate $p_M(d)$ for two typical values of $p$ in Fig.\@ \ref{fig_error} with $p_M(d_0) = p$ to approximate physical injection of $|T\rangle$ and an optimal choice of base code distance $d_0$.
It clearly shows the growing gap between surface code and magic state errors with increasing code distance.
It is possible to close this gap by using a code distance at the top level of the recursive distillation process that is higher than the output code distance.
Using a lower bound from Eq.\@ (\ref{amp_bound}), the error rates are asymptotically balanced by the choice
\begin{equation} \label{code_balance}
 d_{\mathrm{out}} \sim \left( 1 + \frac{ \log \left[ 450 p \Omega(d')^2\right]}{d' \log(143 p)} \right) d_{\mathrm{in}}
\end{equation}
 where $d_{\mathrm{in}}$ is the top-level code distance of the recursive distillation process and $d_{\mathrm{out}}$ is the surface code distance used to store the $|T\rangle$ output and perform the rest of the quantum computation.
Thus a growing error amplification can be mitigated by a multiplicative rescaling of the code distance used for distillation if the proportionality constant is positive in Eq.\@ (\ref{code_balance}).
For example, $p = 7 \times 10^{-4}$, $d' = 7$, and $\Omega(d') = 100$ result in $d_{\mathrm{out}} \sim 0.5 d_{\mathrm{in}}$.
Physical injection of $|T\rangle$ limits the attainable values of $\Omega(d_0)$, but other methods such as magic state cultivation \cite{cultivation} can further reduce $\Omega(d_0)$ and the need to rescale the code distance.

\begin{figure}
\includegraphics{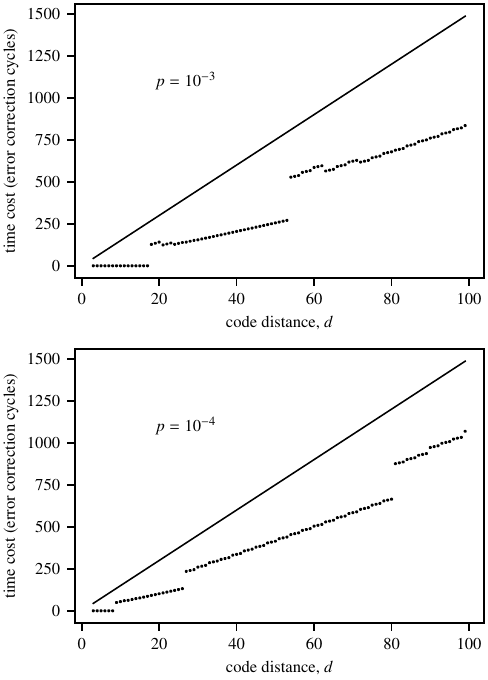}
\caption{Comparison between a numerical estimate (points) and a nominal value of $15 d$ (solid line) for the time cost of the recursive multi-level distillation operations shown in Fig.\@ \ref{fig_error}.
}\label{fig_time}
\end{figure}

Finally, I analyze the validity of the simple time cost estimate in Eq.\@ (\ref{multi_cost_bound}) for the multi-level $|T\rangle$ distillation process.
It both overestimates cost because it assumes infinite levels of distillation and underestimates cost because it does not account for distillation failures.
In Fig.\@ \ref{fig_time}, I numerically estimate the time costs associated with the distillation operations shown in Fig.\@ \ref{fig_error}.
I recursively divide the time cost of each distillation level by
\begin{equation}
 1 - 15[ p_M(\lfloor d/3 \rfloor) + 67.5 p_L(\lfloor d/3 \rfloor) ]
\end{equation}
 at an output code distance of $d$, which is a leading-order estimate of the expected number of repetitions for a successful outcome at that level of distillation consistent with Eq.\@ (\ref{propagate_error}).
It is apparent that the overestimation exceeds the underestimation in the nominal cost of $15d$.

\section{Discussion\label{discussion_section}}
\vspace*{-0.25cm}

While this paper has focused exclusively on magic state distillation, there are several other fundamentally different approaches to logical non-Clifford gate design on topological codes.
Topological codes in three spatial dimensions have transversal non-Clifford gates that are compatible with code switching to planar topological codes \cite{3d_topological_code}.
These constructions can be adapted to implement logical non-Clifford gates directly on planar topological codes by replacing one spatial dimension with time \cite{topological_nonclifford}.
These logical non-Clifford gates have further been generalized to lattice surgery operations between Abelian and non-Abelian planar topological codes \cite{nonabelian1,nonabelian2}.
Another distinct approach to logical non-Clifford gate design is pieceable fault tolerance \cite{pieceable}, which decomposes a logical gate into pieces that can be interleaved with error correction cycles to detect errors and correct them before they spread.
Pieceable fault tolerance was introduced for low-distance codes, but it has also been studied for high-distance codes that are related to planar topological codes \cite{pieceable2}.

All of these alternatives to magic state distillation are appealing because they directly operate on high-distance codes, whereas magic state distillation networks require codes over a wide range of distances.
They also enable non-Clifford gates that pack more neatly into a prescribed spacetime volume with overall costs that are competitive with magic state distillation at low physical error rates \cite{comparison}.
However, none of these alternatives are yet compatible with the planar surface code on a square grid of data qubits and instead require more complicated topological codes with significantly lower physical error thresholds.
The recursive magic state distillation design presented in this paper also packs tightly into a prescribed spacetime volume, and there is perhaps a deeper significance or common cause to the lowered error threshold that results.
All of these approaches to logical non-Clifford gate design have the same ultimate goal of minimizing spacetime cost on a lattice of physical qubits with local gate connectivity.
Magic state distillation can manage its threshold problem by increasing the distance of the parent surface code to reduce the error probability of the output magic states, and topological non-Clifford gate designs can manage their threshold problems by concatenation with a high-threshold base code.
These mitigation strategies increase the spacetime cost of the non-Clifford gate, but expand the domain of applicability of the various design approaches.

\begin{figure}
\includegraphics{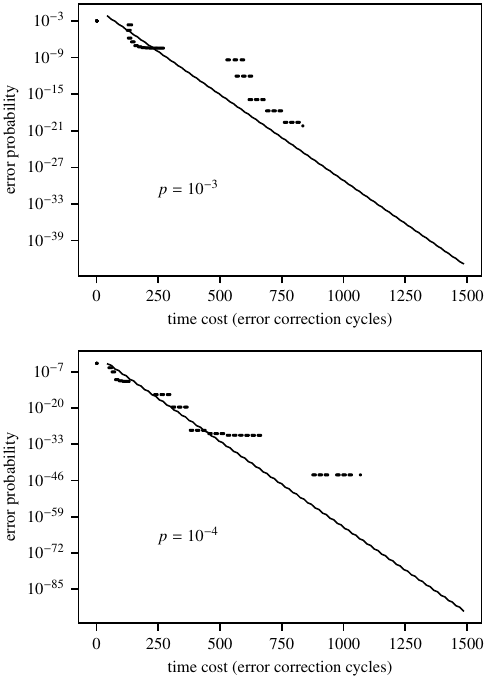}
\caption{Correlation between the error probabilities in Fig.\@ \ref{fig_error} and the time costs in Fig.\@ \ref{fig_time}.
The increase in error probability is partly compensated by the reduction in time cost relative to the nominal behavior.
The granularity of the distillation process limits the accessible regions of cost and error.
}\label{fig_merge}
\end{figure}

While the designs in this paper followed a simple, rigid packing scheme, there are also benefits to more flexible layouts of magic state distillation networks.
Rather than have all levels of distillation occur in the same computational region, successive levels can be placed in neighboring computational regions and magic states can be moved between regions during the distillation process.
Most existing magic state factory designs follow this approach, and it provides more flexibility to adjust the surface code distance independently for each level of distillation.
An increase in space costs from increasing the code distance can be partly mitigated by the early termination of the distillation process because of its granularity, as shown in Fig.\@ \ref{fig_merge}.
Another form of flexibility that is worthwhile to pursue is distillation circuits adapted to non-uniform errors in input states.
Rather than losing distance and wasting space when the code distance is not evenly divisible between levels of distillation, it may be possible to distill inputs on codes with distances that vary by one and fully preserve the overall distance.
This would require a careful alignment of logical error patterns to limit the number of error locations that use a lower-distance code, which is yet another constraint on the design space.

This paper focuses on implementations of well-established many-to-one distillation protocols, but there are benefits to considering a wider range of protocols and searching for new protocols.
Block distillation protocols with multiple magic state outputs \cite{block1,block2} can be more efficient than the single-output protocols considered here, but they require the separation of outputs to avoid correlated errors in multi-level distillation.
Correlated errors are not a problem at the top level of distillation, so block distillation protocols may be well suited as more efficient terminal distillation steps.
However, a block distillation protocol with a large ratio of output to input states might not be more efficient unless it also can be implemented with a competitive spacetime cost.
Among new distillation protocols, an appealing target is quadratic 4-to-1 distillation, which would have a much simpler and more efficient recursive structure than cubic 15-to-1 distillation.
Quadratic 4-to-1 distillation is possible for qudits of prime dimension greater than three \cite{qudits}, but no such protocol has been found for qubits.
It would be worthwhile to search for such protocols more exhaustively, considering more complicated magic states \cite{synth} and the non-Abelian stabilizer group structure used in this paper.

\section{Conclusion}
\vspace*{-0.25cm}

The main goal of this paper has been to reduce the substantial difference in cost between logical Clifford gates and logical non-Clifford gates on the surface code.
These results are summarized in Table \ref{table_cost} with similar space costs between these two types of gates and a range of time costs that vary by less than a factor of ten.
While promising, these resource estimates are an oversimplification that do not account for an amplification of error probabilities in magic states that occurs at typical physical error rates, which I analyzed in Sec.\@ \ref{analysis_section}.
Even so, the recursive magic state distillation design developed in this paper enables universal quantum computation on one surface code tile with only three ancilla tiles needed to implement a single-qubit Clifford+$T$ gate set.
It is easier to run a computer longer than to build a larger computer, thus quantum computing designs that reduce space costs are particularly important.

I also introduced several technical components in this paper that may be useful in other work.
I considered a class of measurement-based distillation circuits and derived their distinct fault tolerance requirements.
I then constructed two examples of such fault-tolerant distillation circuits through both rational design and randomized, computer-aided search to minimize their spacetime cost.
I used expanded non-Abelian stabilizer groups to define logical non-Clifford gates on stabilizer codes without having to evaluate their explicit action on logical states.
Finally, I used a simple error model and asymptotic analysis to estimate the propagation of errors through the recursive multi-level distillation process and identify several different regimes of behavior that depend on the physical error rate.
Further exploration of this design space may uncover more efficient magic state distillation protocols and their recursive implementation on the surface code.

\begin{acknowledgments}
I thank Craig Gidney for helpful discussions about his $S$ gate design.
The Molecular Sciences Software Institute is supported by grant CHE-2136142 from the National Science Foundation.
\end{acknowledgments}

\end{document}